\documentstyle[12pt,epsfig,subfigure]{article}
\catcode`\@=11   
\@addtoreset{equation}{section}

\newcommand{\ccaption}[2]{
    \begin{center}
    \parbox{0.85\textwidth}{
      \caption[#1]{\it \small{{#2}}}}                   
    \end{center}    }
\begin{document}
 
\def    \pt     {\mbox{$p_T$}}
\def\ie{ {\it i.e.} }
\def\eg{ {\it e.g.} }
\newcommand{\be}{\begin{equation}}
\newcommand{\ee}{\end{equation}}
\newcommand{\br}{\begin{eqnarray}}
\newcommand{\er}{\end{eqnarray}}
\newcommand{\ba}{\begin{array}}
\newcommand{\ea}{\end{array}}
\newcommand{\bi}{\begin{itemize}}
\newcommand{\ei}{\end{itemize}}
\newcommand{\bn}{\begin{enumerate}}
\newcommand{\en}{\end{enumerate}}
\newcommand{\bc}{\begin{center}}
\newcommand{\ec}{\end{center}}
\newcommand{\ul}{\underline}
\newcommand{\ol}{\overline}
\newcommand{\eebbww}{$e^+e^-\rightarrow b\bar b W^+W^-$}
\newcommand{\bb}{$ b\bar b \ $}
\newcommand{\ttb}{$ t\bar t \ $}
\newcommand{\ar}{\rightarrow}
\newcommand{\sm}{${\cal {SM}}\ $}
\newcommand{\Dir}{\kern -7.4pt\Big{/}\kern 1.pt}
\newcommand{\Dirin}{\kern -13.4pt\Big{/}\kern 7.4pt}

\newcommand{\DDir}{\kern -7.6pt\Big{/}}
\newcommand{\DGir}{\kern -6.0pt\Big{/}}
\newcommand{\sla}{\kern -5.4pt /}

\newcommand{\dotp}{\!\cdot\!}
\newcommand{\bea}{\begin{eqnarray}}
\newcommand{\eea}{\end{eqnarray}}

\def\Ord{\buildrel{\scriptscriptstyle <}\over{\scriptscriptstyle\sim}}
\def\OOrd{\buildrel{\scriptscriptstyle >}\over{\scriptscriptstyle\sim}}
\def\z0{Z}
\def\gf{G_{\mu}}
\def\zm{M_{_Z}}
\def\bm{m_b}
\def\cm{m_c}
\def\gev{{\hbox{GeV}}}
\def\tev{{\hbox{TeV}}}
\def\nb{{\hbox{nb}}}
\def\msb{{\overline{MS}}}
\def\als{\alpha_{_S}}
\def\tm{m_{t}}
\def\hm{M_{_H}}
\def\wm{M_{_W}}
\def\gn{\Gamma_{\nu}}
\def\ge{\Gamma_{e}}
\def\gmu{\Gamma_{\mu}}
\def\gt{\Gamma_{\tau}}
\def\gl{\Gamma_{l}}
\def\gu{\Gamma_{u}}
\def\gd{\Gamma_{d}}
\def\gc{\Gamma_{c}}
\def\gs{\Gamma_{s}}
\def\gb{\Gamma_{b}}
\def\gz{\Gamma_{_Z}}
\def\gh{\Gamma_{h}}
\def\gi{\Gamma_{\rm {inv}}}
\def\afb{A_{_{\rm {FB}}}}
\def\alr{A_{_{\rm {LR}}}}
\def\gv{g_{_V}}
\def\ga{g_{_A}}
\def\barf{\overline f}
\def\barq{\overline q}
\def\barb{\overline b}
\def\bart{\overline t}
\def\barc{\overline c}
\def\gvf{g^f_{_{V}}}
\def\gaf{g^f_{_{A}}}
\def\gvl{g^l_{_{V}}}
\def\gal{g^l_{_{A}}}
\def\gvb{g^b_{_{V}}}
\def\gab{g^b_{_{A}}}
\def\gsvb{g^b_v}
\def\gsab{g^b_a}
\def\ord {\cal O}
\def\ical{\cal I}
\def\shat{\hat s}
\def\chat{\hat c}
\def\vhat{\hat v}
\def\thetahat{\hat \theta}
\def\alphahat{\hat \alpha}
\def\fvf{F_{_V}^f}
\def\faf{F_{_A}^f}
\def\fvl{F_{_V}^l}
\def\fal{F_{_A}^l}
\def\acal{\cal A}
\def\ste{\sin\theta}
\def\stes{\sin^2\theta_{\rm{eff}}}
\def\xhat{\hat x}
\def\piv{\Pi_{_V}}
\def\pia{\Pi_{_A}}
\def\dr{\Delta r}
\def\drl{\Delta r_{_L}}
\def\dgvf{\delta g_{_V}^f}
\def\dgaf{\delta g_{_A}^f}
\def\dgvl{\delta g_{_V}^l}
\def\dgal{\delta g_{_A}^l}
\def\i3f{I^{(3)}_f}
\def\pih{{\hat\Pi}}
\def\sgh{{\hat\Sigma}}
\def\osp2{16\,\pi^2}
\def\ap2{\left(p^2\right)}
\def\stw{s_{\theta}}
\def\ctw{c_{\theta}}
\def\stws{s_{\theta}^2}
\def\stwf{s_{\theta}^4}
\def\ctws{c_{\theta}^2}
\def\Szg{\Sigma_{_{Z\gamma}}}
\def\Szz{\Sigma_{_{ZZ}}}
\def\Sww{\Sigma_{_{WW}}}
\def\Swwg{\Sigma_{_{WW}}^{^G}}
\def\Stg{\Sigma_{_{3Q}}}
\def\Stt{\Sigma_{_{33}}}
\def\Pgg{\Pi_{\gamma\gamma}}
\def\Pf{\Pi_{_F}}
\def\rhou{\rho_{_U}}
\def\ku{\kappa_{_U}}
\def\rhoz{\rho_{_Z}}
\def\rhozr{\rho^{\scriptscriptstyle R}_{_Z}}
\def\gfd{\gamma_5}
\def\gau{\gamma^{\alpha}}
\def\gad{\gamma_{\alpha}}
\def\mev{{\hbox{MeV}}}
\def\gev{{\hbox{GeV}}}
\def\tmo{\times 10^{-1}}
\def\tmt{\times 10^{-2}}
\def\tmth{\times 10^{-3}}
\def\tmf{\times 10^{-4}}
\def\tmfv{\times 10^{-5}}
\def\srt{\sqrt{2}}
\def\xsf{\sigma_{_F}}
\def\xsb{\sigma_{_B}}
\def\chig{\chi_{\gamma}}
\def\chiz{\chi_{_Z}}
\def\s0h{\sigma^h_0}
\def\ba{\begin{eqnarray}}
\def\ea{\end{eqnarray}}
\def\nl{\nonumber \\}
\def\baral{\bar{\alpha}}
\newcommand{\epem}      {\mbox{$ {\mathrm e}^+ {\mathrm e}^-$}}
\def    \snu            {\mbox{$\tilde{\nu}_e$}}            
\def    \squ            {\mbox{$\tilde{q}$}}
\def    \stop           {\mbox{$\tilde{t}$}}
\def    \stopone        {\mbox{$\tilde{t}_1$}}
\def    \stoptwo        {\mbox{$\tilde{t}_2$}}
\def    \smu            {\mbox{$\tilde{\mu}$}} 
\def    \stau           {\mbox{$\tilde{\tau}$}} 
\def    \sel            {\mbox{$\tilde{e}$}} 
\def    \sele           {\mbox{$\tilde{e}$}} 
\def    \sell           {\mbox{$\tilde{e_L}$}} 
\def    \selr           {\mbox{$\tilde{e_R}$}} 
\def    \mt             {\mbox{$m_t$}}       
\def    \mb             {\mbox{$m_b$}}  
\def    \msnu           {\mbox{$m_{\tilde{\nu}}$}}
\def    \msnue          {\mbox{$m_{\tilde{\nu}_e}$}}
\def    \mslep          {\mbox{$m_{\tilde{\ell}}$}}
\def    \mchipm         {\mbox{$m_{\tilde\chi^\pm}$}}
\def    \mchipma       {\mbox{$m_{\tilde\chi^{\pm}_1}$}}
\def    \mchioa        {\mbox{$m_{\tilde\chi^0_1}$}}
\def    \mchiob        {\mbox{$m_{\tilde\chi^0_2}$}}
\def    \mstop          {\mbox{$m_{\tilde{t}}$}}
\def    \mstopone       {\mbox{$m_{\tilde{t}_1}$}}
\def    \msmu           {\mbox{$m_{\tilde{\mu}}$}}
\def    \msele          {\mbox{$m_{\tilde{e}}$}}
\def    \chipm         {\mbox{$\tilde\chi^{\pm}$}}   
\def    \chipma        {\mbox{$\tilde\chi^{\pm}_1$}}
\def    \chipa         {\mbox{$\tilde\chi^+_1$}}
\def    \chima         {\mbox{$\tilde\chi^-_1$}}
\def    \chipmb        {\mbox{$\tilde\chi^{\pm}_2$}}
\def    \chimpa        {\mbox{$\tilde\chi^{\mp}_1$}}
\def    \chimpb        {\mbox{$\tilde\chi^{\mp}_2$}}
\def    \chio          {\mbox{$\tilde\chi^0$}}
\def    \chioi         {\mbox{$\tilde\chi^0_i$}}
\def    \chioj         {\mbox{$\tilde\chi^0_j$}}
\def    \chioa         {\mbox{$\tilde\chi^0_1$}}
\def    \chiob         {\mbox{$\tilde\chi^0_2$}}
\def    \missEt      {\ifmmode{/\mkern-11mu E_t}\else{${/\mkern-11mu E_t}$}\fi}
\def    \missE          {\ifmmode{/\mkern-11mu E}\else{${/\mkern-11mu E}$}\fi}
\def\te{\tilde e}
\def\tnu{\tilde\nu}
\def\esl{E\llap/}
\def\tg{\tilde g}
\def\tnu{\tilde\nu}
\def\tell{\tilde\ell}
\def\tq{\tilde q}
\def\tb{\tilde b}
\def\tst{\tilde t}
\def\ttau{\tilde\tau}
\def\tl{\tilde\ell}
\def    \foton          {\mbox{$\gamma$}}
\def    \zboson        {\mbox{$Z^{0}$}}
\def    \wboson        {\mbox{$W^{\pm}$}}
\def    \sneutrino     {\mbox{$\tilde{\nu}$}}
\def    \DFGT          {{DFGT }}
\def    \SUSYGEN       {{SUSYGEN }}
\def    \ISAJET        {{ISAJET }}
\def    \mzer          {\mbox{$m_{0}$}}
\def    \Mdue          {\mbox{$M_{2}$}}
\def    \Mu            {\mbox{$\mu$}}
\def    \tanbet        {\mbox{$\tan\beta$}}
\def    \HELAS         {{HELAS }}
\def    \BASES         {{BASES }}
\def    \SPRING        {{SPRING }}
\def    \JETSET        {{JETSET 7.4 }}
\def    \Gcharg        {\mbox{$\Gamma_{\tilde\chi^{\pm}}$}}
\def    \Xsect         {\mbox{$\sigma$}}
\def    \chreaA      {\mbox{$\tilde\chi^{\pm}_1\rightarrow\tilde{\nu}l^{\pm}$}}
\def    \chreaB      {\mbox{$\tilde\chi^{\pm}_1\rightarrow W^{*}\tilde\chi^{0}$}}
\def    \chreaC      {\mbox{$\tilde\chi^{\pm}_1\rightarrow\tilde{f}^{*}\nu$}}
\def    \Mchar         {\mbox{$m_{\tilde\chi^{\pm}}$}}                  
\def    \Mneut         {\mbox{$m_{\tilde\chi^{0}}$}}
\def    \Mslep         {\mbox{$m_{\tilde{l}}$}}
\def    \Msquar        {\mbox{$m_{\tilde{q}}$}}
\def    \Msneut        {\mbox{$m_{\tilde{\nu}}$}}
\def    \as             {\mbox{$\alpha_{\rm s}$}}     
\def    \ret            {\\[\eqskip]}
\def    \to             {\rightarrow }
\def    \as             {\mbox{$\alpha_{\rm s}$}}     
\def\ti              {\tilde}
\def    \stop           {\mbox{$\tilde{t}$}}

\def\stst            {\stop \bar{\stop}}
\def\sb              {\ti b}
\newcommand {\ra} {\rightarrow}
\newcommand {\bao} {\overline}
\newcommand {\LQ} {{\sc Lq2}}
\def\tanb{\mbox{$\tan\beta$}}
\def\Gcs{\hbox{GeV}/\mbox{$c^2$}}
\def\Tcs{\hbox{TeV}/\mbox{$c^2$}}
\def\epemto{\mbox{$\hbox{e}^+\hbox{e}^- \to$}}
\def\epem{\mbox{$\hbox{e}^+\hbox{e}^-$}}
\def\Z{\mbox{$\hbox{Z}$}}
\def\W{\mbox{$\hbox{W}$}}
\def\A{\mbox{$\hbox{A}$}}
\def\H{\mbox{$\hbox{H}$}}
\def\h{\mbox{$\hbox{h}$}}
\def\g{\mbox{$\hbox{g}$}}
\def\mpmm{\mbox{$\mu^+\mu^-$}}
\def\tptm{\mbox{$\tau^+\tau^-$}}
\def\lplm{\mbox{$\hbox{l}^+\hbox{l}^-$}}
\def\nnbar{\mbox{$\nu\bar\nu$}}
\def\ffbar{\mbox{$\hbox{f}\bar{\hbox{f}}$}}
\def\qqbar{\mbox{$\hbox{q}\bar{\hbox{q}}$}}
\def\ssbar{\mbox{$\hbox{s}\bar{\hbox{s}}$}}
\def\ccbar{\mbox{$\hbox{c}\bar{\hbox{c}}$}}
\def\bbbar{\mbox{$\hbox{b}\bar{\hbox{b}}$}}
\def\ttbar{\mbox{$\hbox{t}\bar{\hbox{t}}$}}
\font\eightrm=cmr8
\def\mh{\mbox{$m_{\hbox{\eightrm h}}$}}
\def\mH{\mbox{$m_{\hbox{\eightrm H}}$}}
\def\mA{\mbox{$m_{\hbox{\eightrm A}}$}}
\def\mZ{\mbox{$m_{\hbox{\eightrm Z}}$}}
\def\mW{\mbox{$m_{\hbox{\eightrm W}}$}}
\def\mSUSY{\mbox{${\hbox{M}}_{\hbox{\eightrm SUSY}}$}}
\def\At{\mbox{${\hbox{A}}_{\hbox{\eightrm t}}$}}
\def\Ab{\mbox{${\hbox{A}}_{\hbox{\eightrm b}}$}}
{\flushright{
        \begin{minipage}{6cm}
        hep-ph/9602203\hfill \\
        \end{minipage}        }
}
\begin{center}
{\large \bf EVENT GENERATORS FOR DISCOVERY PHYSICS}
\end{center}                                       
\begin{center}
{\it Conveners}: M.L.~Mangano and G.~Ridolfi
\end{center}          
\begin{quote}
{\it Working group}: E.~Accomando, S.~Asai, H.~Baer, A.~Ballestrero,
M.~Besan\c{c}on,        
E.~Boos, C.~Dionisi, M.~Dubinin, L.~Duflot, V.~Edneral, K.~Fujii, J.~Fujimoto, 
S.~Giagu, D.~Gingrich,
T.~Ishikawa, P.~Janot, M.~Jimbo, T.~Kaneko, K.~Kato, S.~Katsanevas,
S.~Kawabata,  S.~Komamiya, T.~Kon, Y.~Kurihara, A.~Leike, 
G.~Montagna,                                  
O.~Nicrosini, F. Paige, G.~Passarino, D.~Perret-Gallix, 
F.~Piccinini, R.~Pittau, S.~Protopopescu,           
A.~Pukhov, T.~Riemann, S.~Shichanin, Y.~Shimizu, A.~Sopczak,
H.~Tanaka, X.~Tata, T.~Tsukamoto                           
\end{quote}         
\tableofcontents
\newpage
\section{Introduction}
This chapter of the report presents a review of Monte Carlo (MC) event
generators for signals of new particles. The areas
covered include Higgs production, Supersymmetry (SUSY) and leptoquarks.     
Contrary to other contexts, where MC generators for specific Standard Model
(SM) processes are
considered, it is not possible to identify a simple common set of
features which event generators for new physics should possess. Each new
process presents its own theoretical and technical issues, with the emphasis
being now equally shared between the precision of the calculations 
and the completeness of the
coverage of exotic phenomena and their parametrization. While the accuracy and
the statistical power of the future measurements call for high precision in the
Bhabha, $WW$ and QCD generators, a precision of the order of few percent in the
determination of the cross sections for new phenomena and for their backgrounds
is sufficient in most examples of practical relevance. 
In this respect, it is important to distinguish
between two uses of event generators for new physics. The first one
involves the evaluation of the potential signals, \ie the calculation of
production cross sections, decay branching ratios (BR's) and detector
acceptances and efficiencies. The second one involves the determination of the
parameters of the new physics which will be hopefully discovered from the
comparison of the properties of the observed signal with what derived
from the MC model. Most of the studies carried out by our working group
and by the New Physics working groups covered the first issue. In the
examples considered, the conclusion 
was that the current theoretical uncertainties in the various MC's
do not affect the projected discovery potential.
On the other hand
the extraction of the parameters which determine the specific model of new
physics could depend strongly on the accuracy of the theoretical description
of the production process. For example, features such as the presence or
absence of  spin correlations, which do not seem to be critical for the
discovery of supersymmetric particles, will affect the determination of the
EW properties of the new particles, as will be shown explicitly in
sect.~\ref{DFGT}.
                                                                         
The plan of this contribution is as follows: we start with Higgs production,
shortly describing the main technical issues and presenting the available
generators. Results and comparisons are discussed. 
We then present the SUSY generators, covering both multi-purpose codes which
include most of the possible SUSY final states, and single-channel codes, which
focus on a given signal trying to incorporate the most accurate 
theoretical treatment possible today.
The description of a leptoquark generator will complete this work.

While this review is by no means complete, it contains most of the tools
available to the public. We are aware of many other existing programs, part of
which have been used in the extensive cross checks performed as part of the
working group activity. Since they have not been developed for distribution,
and would not be easily accessible to the public, 
they have not been included in this report.

\section{Higgs}                                                     
The search for the Higgs boson will have first priority in the LEP2 programme
\cite{higgssection},
and a large effort has been devoted to the development of reliable MC event
generators. In the Standard Model, Higgs production at LEP2 is dominated by the
process $e^+e^- \to Z^* \to ZH$ \cite{higgssection}. 
In the mass range of interest for LEP2
the Higgs boson is expected to decay dominantly into a pair of bottom
quarks, leading to final states like $f\bar f b \bar b$, $f$ being any fermion
aside from the top. 
Because of the large width of the $Z$ boson, the approximation in which the
production and decay of the $Z$ boson factorize is not good enough. On the
other hand, the small width of the Higgs could justify the factorization
approximation. Nevertheless, most of the 
event generators presented below include the matrix elements for the full
4-fermion process $e^+e^- \to f \bar f b\bar b$. The evaluation of this
process involves not only the diagrams with a Higgs boson, but also
all possible SM diagrams leading to the same final state. As an
example, assuming $f\ne e,\nu_e$ one should
evaluate a total of 25 tree level diagrams: 1 corresponding to the signal, 8 $t$-channel
diagrams relative to $ZZ$, $Z\gamma$ and $\gamma\gamma$ exchange, and 16
$s$-channel diagrams relative to the bremstrahlung of a neutral vector boson
from the fermionic final states. If $f$ is a quark, QCD processes should be
added to this last category. Likewise, different sets of
diagrams appear both in the signal and in the background if $f=e$ or $f=\nu_e$.
The presence of several resonating channels in                                 
the full amplitude poses some numerical problem, which can be easily overcome 
by choosing properly the importance sampling, as described later on.             
In the case of massless final state fermions, the interference
between signal and background diagrams is zero, because of the helicity 
non-conservation induced by the coupling to the Higgs boson. If the mass of the
$b$ quarks is kept different from zero in the matrix elements, a finite
interference will develop. In addition to including all diagrams, accurate
event generators should also include the effects of initial state radiation
(ISR),
and provide the user with the effective 4-momentum of the final state after
initial state photon emission. 
As a desirable feature, Higgs generators should also contain a description of
Higgs production and decay in models beyond the SM, such as two-doublet
or SUSY models \cite{higgssection}.                                     
Finally, one expects the code to provide
unweigthed events with the 4-momenta of all final state particles, in order for
the user to process the events through the detector and to apply analysis cuts.

Each code presented in this section embodies all these features to
a different degree. A comparison between results obtained using different
approximations will allow us to estimate the importance of any given effect,
and to assess the limitation of a given approach.
It must be pointed out that none of these codes contains the full 1-loop EW
radiative corrections. Their evaluation and inclusion in a 4-fermion event 
generator has not
been achieved for any 4-fermion final state.
The largest component of the       
radiative corrections is however incorporated using the so-called Improved
Born Approximation \cite{IBA}, in which vector boson self-energy insertions are
absorbed by using running EW couplings. A partial calculation of the full EW
1-loop corrections has been performed \cite{Gross} for the process
$e^+e^- \to H f \bar f$. The resulting production cross section never differs
from the IBA by more than 2\% in the range of interest at LEP2. 
The agreement improves for Higgs masses near the LEP2 discovery reach. 
The 1\% level is therefore an optimal goal for the agreement between
the tree level event generators which will be described here.

\subsection{CompHEP}
\begin{center}                             
\begin{tabular}{ll}
Program name:          & CompHEP -- version 3.0 \\
Authors:               & E. Boos  -- {\tt boos@theory.npi.msu.su} \\
                       & M. Dubinin -- {\tt dubinin@theory.npi.msu.su} \\
                       & V. Edneral -- {\tt edneral@theory.npi.msu.su} \\
                       & V. Ilyin -- {\tt ilyin@theory.npi.msu.su} \\
                       & A. Pukhov -- {\tt pukhov@theory.npi.msu.su} \\
                       & V. Savrin -- {\tt savrin@theory.npi.msu.su} \\
                       & S. Shichanin -- {\tt shichanin@m9.ihep.su} \\
Availability:          & anonymous ftp from {\tt theory.npi.msu.su}\\
                       & Directory: pub/comphep-3.0\\              
                       & File:  {\tt 30.tar.Z}\\
Documentation:         & Files: {\tt install.doc, manual.ps.Z}\\
\end{tabular}
\end{center}

The main purpose of CompHEP \cite{r1} is to allow the automatic evaluation of
cross sections and distributions directly from an assigned lagrangian. 

The general structure of the CompHEP package is described in the
section "Event generators for WW physics" of this Workshop.
Here we describe in more detail the feature of the program relevant for
the Standard Model (SM) Higgs search at LEP2.

Any kind of three-, four- and five-particle final states can be calculated
using CompHEP. In the case of Higgs boson production, the reactions of interest
are $\epem \to f \bar f b \bar b$, with $f$ any lepton or quark.
The main features of the
calculations implemented in CompHEP
can be summarized in the following way:
\begin{itemize}
\item[--] all possible Feynman diagrams contributing to the process are
calculated and all interferences between signal and background diagrams are
taken into account (at tree level). Fermion masses can be kept nonzero in the
calculation of the squared amplitudes.
\item[--] final particle phase space with massive fermions is generated
explicitly. 
\end{itemize}         
CompHEP generates graphically complete sets of Feynman diagrams for the
processes mentioned above (for instance, 25 diagrams including one signal
diagram for $e^+ e^- \rightarrow \mu^+ \mu^- b \bar b$, 21 diagrams including
two signal diagrams for $e^+ e^- \rightarrow \nu \bar \nu b \bar b$ , 50
diagrams including two signal diagrams for $e^+ e^- \rightarrow e^+ e^- b \bar
b$). Any desired subset of diagrams (for instance, signal only) can be
separated for further processing. Squared amplitudes and interference terms are
calculated symbolically with the help of a special module for trace
calculations. In the next step, optimized FORTRAN codes corresponding to these
terms are generated by the package. The codes are compiled and linked to the
special interface program and Monte Carlo integrator program. The
FORTRAN loading module created as a result of this process represents by itself
the generator of the Higgs signal in the four fermion reaction under
consideration. It is driven by the screen menu allowing the user to choose
various options of signal and background simulation. A more detailed
description of the menu system can be found in ref.~\cite{wwsection}.
                                                                     
Seven-dimensional adaptive Monte Carlo integration over the phase space and
unweighted event generation is performed by the 
BASES/SPRING package \cite{bases}.
The output has the standard BASES form (sequence of Monte Carlo iterations for
total cross section and a set of histograms for various distributions). The
width of the light Higgs boson is small, so the 
adaptive possibilities of BASES are
not sufficient for integration over the phase space. Additional kinematical
regularization (integration with probability density concentrated around the
resonance peaks) can be introduced for the Higgs as well as vector bosons. 

Initial state radiation is implemented in the structure function
approach. Non-standard interaction vertices can be introduced by
changing the model input (see \cite{r1} for details). Any kinematical
cuts can be implemented.

At present, versions of CompHEP for different platforms
exist: HP Apollo 9000, 
IBM RS 6000, DECstation 3000, SPARC station, Silicon Graphics
and VAX.
\subsection{{\tt 4fan}}
\begin{center}                             
\begin{tabular}{ll}
Program name:          & {\tt 4fan}\\
Authors:               & D.~Bardin -- {\tt bardindy@cernvm.cern.ch} \\
                       & A.~Leike -- {\tt leike@cernvm.cern.ch} \\
                       & T.~Riemann -- {\tt riemann@ifh.de} \\
Availability:          & Anonymous ftp from
                         {\tt gluon.hep.physik.uni-muenchen.de:4fan.}\\
                       & Files: {\tt 4fanv12.f, 4fanv12.dat, readme}\\
Documentation:         & D. Bardin, A. Leike and T. Riemann,   
                         Phys. Lett. {\bf B344} (1995) 383,  \\             
                       & D. Bardin, A. Leike and T. Riemann,   
                         Phys. Lett. {\bf B353} (1995) 513.                   
\end{tabular}
\end{center}

{\tt 4fan} is a semi-analytical program which calculates the process
\begin{equation}\nonumber
e^+e^-\rightarrow f_1\bar f_1 f_2\bar f_2,
\end{equation}
where the three involved fermions $e,f_1$ and $f_2$ must be in different
electroweak multiplets (the so called NC32 process) \cite{nc24}.
SM
Higgs production can be included optionally \cite{Higgs_g}.
For calculations at the Born level, {\tt 4fan} can be used as a
stand-alone program.
For the calculation of cross sections including initial state radiation, the
initial state radiation environment of the code {\tt GENTLE} has to be used
which calls {\tt 4fan} as a subroutine. For the description of {\tt
GENTLE/4fan},  we refer to~\cite{wwsection}.  Here we describe the stand-alone
program {\tt 4fan}.
                                                                        
Six of the eight integrations of the four particle phase space
were done
analytically. The two remaining integrations over
$s_1=[p(f_1)+p(\bar f_1)]^2$ and $s_2=[p(f_2)+p(\bar f_2)]^2$
are performed numerically allowing the inclusion of cuts for these variables.

Finite mass effects are taken into account using the following
approximations:
\begin{description}
\item The phase space is treated exactly.
\item In the Higgs contributions and the conversion diagrams
$e^+e^-\rightarrow (\gamma\gamma)\rightarrow f_1\bar f_1 f_2\bar f_2$,
the masses are treated up to order $O[m^2(f_i)/s_i]$.                          
\item Fermion masses are treated identically in traces and Higgs couplings.
\item The Higgs width is calculated including the decays into $b$-, $c$-
and $\tau$- pairs.
\end{description}
The numbers quoted in the tables of sect.~\ref{higgscomp}
are produced for zero fermion masses except in the Higgs couplings.
The Higgs propagator is always connected with $s_2$ by convention.

The initialization routine {\tt BBMMIN} assigns to the SM parameters 
the values from the Particle Data Book \cite{PDG}.
In the subroutine {\tt DSDSHSZ}, the interferences between the three main
subsets of the NC32 diagrams are calculated as well as those with the
Higgs signal diagram. Their sum gives the double differential cross section.
The integration of selected interferences between these subsets is not
foreseen.

The numerical integration is done by a twofold application of a one-dimensional
Simpson integration with a control over the relative and the absolute error. The
singularities due to resonating vector propagators are eliminated by appropriate
changes of integration variables. To avoid numerical instabilities, the
kinematical functions resulting from the six-fold analytical integration are
replaced by Taylor expansions near the borders of the phase space. The shortest
calculational time is achieved by a choice of the required absolute and relative
errors in such a way that they give approximately equal contributions to the
error of the output. 

The calculational time of a Born cross section is several seconds on a HP
workstation depending on the required accuracy and on the cuts on $s_1$ and
$s_2$; improving the accuracy by a factor of ten approximately doubles the
calculational time. 

Input and output are transferred through the arguments of the subroutine only.

Usage of the program:\\[0.2cm]
{\tt\hspace{6ex}              
CALL FOURFAN(EPS,ABS,IF1,IF2,S,S1MIN,S1MAX,S2MIN,S2MAX,AMH,IOUT,OUT)}

\noindent{\bf Input:}
    
\begin{tabular}{lll}
{\tt EPS,ABS}:&& The required relative and absolute error. If at least
                one of\\ &&
                the two criteria is fulfilled, the calculation
                is finished. \\
{\tt IF1,IF2:}&& Integers specifying the two final fermion pairs
                according \\ && to the Monte Carlo particle numbering scheme,\\
             && see Particle Data Group \cite{PDG}, Chapter 32).\\
{\tt S:} && The c.m. energy squared of the $e^+e^-$ pair.\\
{\tt S1MIN,S1MAX:} && The integration bounds of $s_1$.\\
{\tt S2MIN,S2MAX:} && The integration bounds of $s_2$.\\
{\tt AMH:} && The Higgs mass.\\
{\tt IOUT:}  && Integer, selecting the output, \\ &&
                Currently {\tt IOUT}=1, 2, 11 and 12 are implemented:\\
        & {\tt IOUT=1:} & Total cross section $\sigma_t$ without Higgs.\\
        & {\tt IOUT=2:} & Differential cross section
                          ${\rm d}\sigma /{\rm d} s_2$ without Higgs.\\
        & {\tt IOUT=11, 12:} & The same as {\tt IOUT=1, 2}
                             but {\it with} Higgs.
\end{tabular}\vspace{0.3cm}
The units of the input (if required) are GeV$^2$ or GeV.\vspace{0.3cm}

\noindent{\bf Output:} {\tt OUT}\ \ \
Depends on the value of {\tt IOUT}.
The output is given in $fb$ or in $fb/$GeV.

On HP workstations {\tt 4fan} must be compiled with the -K option.
\\
\subsection{HIGGSPV}
\begin{center}                             
\begin{tabular}{ll}
Program name:          & HIGGSPV \\
Authors:               & G. Montagna -- {\tt montagna@pv.infn.it} \\
                       & O. Nicrosini -- {\tt nicrosini@vxcern.cern.ch} \\
                       & F.~Piccinini -- {\tt piccinini@pv.infn.it} \\
Availability:          & Code available upon request\\
Documentation:         & \\
\end{tabular}
\end{center}

\noindent{\bf General Description.}
The present version of the four-fermion Monte Carlo code HIGGSPV is an upgrade
of the version used in~\cite{hl}, where a general description of the formalism
adopted and the physical ideas behind it can be found (see also references
therein). All the physical and technical upgrades will be described in detail
in~\cite{hcpc}. 

The program is based on the exact tree-level calculation of several
four-fermion final states relevant for Higgs search at future $e^+ e^-$
colliders. Any cut on the final state configuration can be implemented.
Initial- and final-state QED corrections are taken into account at the leading
logarithmic level by proper structure functions, including $p_T / p_L$ 
effects~\cite{cpcww}.                                                         
An hadronization interface is under development. All the relevant presently
known non-QED corrections are also taken into account. 

\noindent{\bf Features of the program.}
The code consists of three Monte Carlo branches, in wich the
importance-sampling technique is employed to take care of the peaking behaviour
of the integrand: 
\begin{itemize}
\item Unweighted event generation. 
The code provides a sample of unweighted events, defined as the components of
the four final-state fermions momenta, plus 
the components of the initial- and final-state photons,
plus $\sqrt {s}$, stored into proper
$n$-tuples. The code returns also the value  of the unweigthed-event cross
section, together with a Monte Carlo estimate of the error. The program must be
linked to CERNLIB for graphical interfaces.
\item Weighted event integration. It is intended for 
computation only. In particular, the code returns the value of the 
cross section for weighted events together with a Monte Carlo estimate 
of the errors. The 
program must be linked to CERNLIB for the evaluation of few special 
functions.
\item Adaptive integration. It is intended for 
computation 
only, but offering high precision performances. On top of importance 
sampling, an adaptive Monte Carlo integration algorithm is used. 
The program must be linked to NAG library for the Monte Carlo adaptive
routines. Full consistency between non-adaptive and adaptive integrations has
been explicitely proven. Neither final-state radiation nor $p_T$ splitting are
taken into account in this branch. 
\end{itemize}
The non-adaptive branches rely upon the random number generator RANLUX. 

The most important features are:
\begin{itemize}             
\item The processes available are the neutral current reactions
$e^+ e^- \to l \bar l q \bar q$, namely 
NC48 (NC50 = NC48 + Higgs signals) NC24 (NC25 =
NC24 + Higgs signal), NC19 (NC21 = NC19 + Higgs signals).
\item Any kind of cuts can be imposed.
\item There is the possibility of getting
information on the contribution of subsets of the diagrams by setting
proper flags.
\end{itemize}
At present, final state decays are not implemented and finite fermion mass
effects are partially taken into account at the phase space boundary. However
it is worth  noting that the ${\cal O}
(\alpha_s^2)$ running quark masses ($m_{c,b} (m_H^2)$)
are employed in the $H q \bar q$ coupling. An interface to hadronization
packages is presently under development. 
     
\noindent {\bf How the code works.}
After the initialization of the SM parameters and of the
electromagnetic quantities, the independent variables are generated, according
to proper multi-channel importance samplings, within the allowed phase space.
By means of the solution of the exact kinematics, the four-momenta of the
outgoing fermions, together with the four-momenta of all the generated photons,
 are reconstructed in the laboratory frame. If the event satisfies the cuts
imposed by the user in SUBROUTINE CUTUSER the matrix element is called,
otherwise it is set to zero. 

In the generation branch, an additional random number is generated in order to
implement the hit-or-miss algorithm and if the event is accepted it is recorded
into an $n$-tuple. 
\noindent
In the non-adaptive integration branch, the cross section for weighted events
is computed. 
In the adaptive integration branch (ref.: NAG routine D01GBF), on top
of importance sampling the
integration routine automatically subdivides the integration region
into subregions  and iterates the procedure where the integrand is
found more variant. The program
stops when a required relative precision is achieved.
    
\noindent {\bf Input parameters and flags.}
A sample of input flags that can be used is  the following: 

\noindent
{\tt OGEN = I} choice between integration [I] and generation [G] 
branch

\noindent
{\tt RS = } c.m. energy (GeV)

\noindent
{\tt OFAST = N} choice between adaptive [Y] or non adaptive [N] branch

\noindent
{\tt NHITWMAX = } number of weighted events

\noindent
{\tt IQED = 1} choice fo Born [0] or QED corrected [1] predictions

\noindent
{\tt OSIGN = Y} includes [Y] or does not include [N] the Higgs-boson signal
                                                 
\noindent
{\tt OBACK = Y} includes [Y] or does not include [N] the SM background
                                                 
\noindent
{\tt NSCH = 2} Renormalization Scheme choice (three possible choices)

\noindent
{\tt ALPHM1 = 128.07D0} $1/\alpha$ value (LEP2 standard input)

\noindent
{\tt ANH = } the Higgs-boson mass (GeV)

\noindent
{\tt OBS = 1} option for the required $l \bar l q \bar q $ channel

\noindent
The Higgs-boson width is calculated including the decays into $c$, 
$\tau$ and $b$ pairs. 
A detailed account of the other relevant possibilities offered by the 
code (namely, command files for generation and adaptive integration 
branches) will be given elsewhere~\cite{hcpc}. 

\noindent {\bf Description of the output.}
For all three branches the output contains the values of the relevant Standard
Model parameters.  In the generation branch, an $n$-tuple containing
the generated events is written,  in addition to the output file containing the
values of the cross sections for unweighted events.
In the integration branches, the values of the cross sections with 
their numerical errors are printed.                           
\subsection{HZHA}
\begin{center}                             
\begin{tabular}{ll}
Program name:          & HZHA \\
Author:                & P. Janot -- {\tt janot@cernvm.cern.ch} \\
Availability:          & {\tt JANOT 193} minidisk on {\tt CERNVM}.\\
                       & Files {\tt HZHA FORTRAN}, {\tt HZHA CARDS} and
                        {\tt HZHA EXEC}. \\
Documentation:         & \\                                                     
\end{tabular}
\end{center}

\noindent{\bf General description.}     
This generator is designed to provide a complete coverage of possible
production and decay channels of 
SM (h) and MSSM (h, H, A) Higgs bosons 
at \epem\ colliders. The complete set of background four-fermion processes is
however not included.                                                        
HZHA allows eight different Higgs production processes to be simulated
(only the processes 1, 5 and 7 are relevant for the SM):
\begin{enumerate}                                                  
        \item  $\epemto\ \h\Z \to \h\ffbar$,
        \item  $\epemto\ \H\Z \to \H\ffbar$,
        \item  $\epemto\ \h\A$,
        \item  $\epemto\ \H\A$,
        \item  $\epemto\ \nnbar\h$ {\it via} WW fusion,
        \item  $\epemto\ \nnbar\H$ {\it via} WW fusion,
        \item  $\epemto\ \epem\h$ {\it via} ZZ fusion,
        \item  $\epemto\ \epem\H$ {\it via} ZZ fusion,
\end{enumerate}
No interference between these channels is as yet included.
The following decay modes of each Higgs boson are 
considered:
\begin{center}
\begin{tabular}{llll}
1.  $\gamma\gamma$  &  2.  $\g\g$  &  3.  $\tptm$   &  4.  $\ccbar$ 
  \\ 
5.  $\bbbar$ &  6.  $\ttbar$  &  7.  $\W^{+\ast}\W^{-\ast}$  &  
                                                        8.  $\Z^\ast\Z^\ast$
  \\ 
9.  $\h, \H \to \A\A$, with $\A\to\Z\h$   &  10.  $\H  \to \h\h$
               &  11.  $\gamma\Z^\ast$   &  12.  $\epem$   \\
13.  $\mpmm$  & 14.  $\ssbar$  &  15.  $\tilde\chi\tilde\chi$  &  16.  $\tilde\chi^+\tilde\chi^-$.      
\end{tabular}                                       
\end{center}
The squark, slepton, chargino, neutralino masses and mixings are computed
in the MSSM framework. The squarks and sleptons are assumed to be sufficiently 
heavy that no Higgs boson can decay to them.
However, decays to \chio's and \chipm's are enabled when
kinematically allowed.                  
Therefore, the branching ratios of charginos and neutralinos are also
computed and their decays simulated in the following channels :
\begin{enumerate}
        \item $\chiob \to  \chioa Z^\ast
          \to \chioa\ffbar$, 
        \item $\chiob \to \tilde\chi^+ \W^{-\ast} \to
          \chioa\ffbar^\prime$,  
        \item $\chiob\to \chioa \gamma$,
        \item $\tilde\chi^+ \to \chioa \W^\ast \to \chioa\ffbar^\prime$,
\end{enumerate}
where \chioa\ and \chiob\ are two lightest neutralinos, and $\tilde\chi^+$ is
the lightest chargino (with $m_{\chiob}, m_{\tilde\chi^+} > m_{\chioa}$).
Cascade decays are also simulated. The lightest neutralino \chioa\ is assumed to
be the LSP (if not so, a warning message appears and the program may stop) and
R-parity is assumed to be conserved. 

In the SM the $\h\to\tilde\chi\tilde\chi$ is allowed to simulate invisible Higgs
decays. In the MSSM, the $\gamma\gamma$ and $\gamma\Z$ (resp. gg) decay widths
are computed with all the charged (resp. coloured) particles in the loops
(squarks, leptons, charginos, charged Higgses). 

Finally, the MSSM Higgs boson pole masses are computed using by default the
improved renormalization group equations at two loops \cite{carena} (they may
also be computed using the EPA for comparison purposes). An independent
computation of Higgs masses \cite{haber} will be implemented soon. 

\noindent{\bf Features of the program.} 
{\tt HZHA} is an event generator based on a Monte-Carlo technique, producing
any desired combination of the final states listed above. In addition, any Z
decay channel combination can be defined by the users for the processes 1 and 2
(\epemto\ \h\Z\ and \H\Z). 

The initial state radiation (ISR) is implemented by means of the REMT package
by R. Kleiss, modified to account for the $\alpha^2$ part of the spectrum, and
the possibility of the radiation of two initial photons. The final state
radiation (FSR) is implemented for the leptonic Z decays in the processes 1 and
2. 

All final state fermions are massive. The couplings of the Higgs bosons to the
quarks are computed using the two-loop running quark masses evolved to the 
Higgs boson mass scale. The pole masses chosen for the c- and the b-quark are
1.64 and 4.87~\Gcs. 
More generally, the cross-sections for all requested processes and the decay
widths/bran\-ching ratios for all three Higgs bosons are computed with all
known QED, weak and QCD corrections. In particular, Higgs width effects
are taken into account both in the cross-section computation and in the
event generation. Finally, the program is fully interfaced with
{\tt JETSET 7.4} \cite{jetset}
for the hadronization of the final state quarks.
                                                
\noindent{\bf How it works.} 
When the program is called, the initialization part determines
the relevant masses, mixing and couplings as mentioned above,
computes the decay widths and branching ratios for the Higgs bosons,
the neutralinos and the charginos, and gives the total production
cross-sections without and with ISR.

Unweighted events are generated according to the user requests
(number of events to be generated, choice of the production processes
and the decay channels,...).
The events are stored in the {\tt LUJETS}
common blocks for subsequent use, e.g. in an interface with full detector
simulation.

The job is closed by some statistics printout (numbers of events
generated in each of the processes and of the decay channels).

\noindent {\bf Input parameters, flags, etc.}
The inputs are chosen by the user through data cards read by the
{\tt CERNLIB} routine {\tt FFREAD}. See item 8 to see where the card
file can be obtained from. This card file is well documented and self
explanatory. The following inputs can be freely set:
\begin{enumerate}
\item From the card {\tt TRIG}, the first and last events to be generated;
\item From the card {\tt DEBU}, the first and last events to be printed out;
\item From the card {\tt TIME}, the time to keep at the end of the job;
\item From the card {\tt GENE}, the general parameters (centre-of-mass energy,
ISR flag, SM or MSSM flag...);
\item From the card {\tt GSMO}, the SM parameters (Z mass and
width, Fermi constant, top mass, Higgs boson mass \mH, $\Lambda_{QCD}^{(5)}$);
\item From the card {\tt GSUS}, the MSSM parameters (\mA, \tanb, the universal
gaugino mass M, the squark mixing parameters $\mu$, \At, \Ab,
and the masses $m_Q$, $m_U$, $m_D$, $m_L$, $m_E$);
\item From the card {\tt PRYN}, the process(es) to be generated;
\item From the card {\tt GZDC}, the Z decay channels to be enabled;
\item From the card {\tt GCH1}, the H decay channels to be enabled;
\item From the card {\tt GCH2}, the h decay channels to be enabled (also
used for the SM Higgs boson);
\item From the card {\tt GCH3}, the A decay channels to be enabled;
\end{enumerate}
Other data cards can be added (in which case the program should be modified
to be able to understand them) to set branching ratios, masses, widths
of particles for the {\tt JETSET} running.

\noindent {\bf A description of the output to be expected.} 
The output contains the values of the Higgs boson
production cross-sections and decay branching ratios, followed by the listing
of the numbers of events given by the data card {\tt DEBU}, and terminated
by the end-of-run statistics.

\noindent{\bf Where can the program be obtained?}
The program (and its subsequent updates) can be obtained upon request
by e-mail to janot@cernvm.cern.ch. It can also be found, for the time
being, on the JANOT 193 minidisk on CERNVM. Relevant files are named
{\tt HZHA FORTRAN} and {\tt HZHA CARDS}. An example of EXEC file
({\tt HZHA EXEC}) can also be found at the same place.
\subsection{PYTHIA}
\begin{center}                             
\begin{tabular}{ll}
Program name:          & {\sc Pythia} -- version 5.720, 29 November 1995 \\
Author:                & T. Sj\"ostrand -- {\tt torbjorn@thep.lu.se} \\
Availability:          & http://thep.lu.se/tf2/staff/torbjorn/Welcome.html \\
Documentation:         & address above and Comp. Phys. Commun. 82 (1994) 74 \\
\end{tabular}
\end{center}
PYTHIA \cite{jetset} is a general-purpose event generator, with emphasis on a 
complete description of QCD cascades and hadronization. Therefore it 
is extensively discussed in the QCD generators report of this 
report. Among its selection of subprocesses, described there,
there are several related to Higgs production (signal and backgrounds).
Extensive details can be found in the documentation referred to above.
\subsection{WPHACT}
\begin{center}                             
\begin{tabular}{ll}
Program name:          & WPHACT \\
Authors:               & E. Accomando -- {\tt accomando@to.infn.it} \\
                       & A.~Ballestrero -- {\tt ballestrero@to.infn.it} \\
Availability:          & Anonymous {\tt ftp} from:\\
                       & {\tt ftp.to.infn.it:pub/ballestrero} \\
Documentation:         & To be found in the above directory.\\                       
\end{tabular}                                            
\end{center}

\noindent {\bf General description.} WPHACT is a program created to study
four fermion, WW and Higgs  physics 
at present and  future $e^+ e^-$ colliders. In its present form, it can 
compute all SM processes with four fermions in the final state. 
\par  We will give here  a description of the program and his characteristics 
with particular emphasis  to those regarding Higgs physics. 
We refer to the analogous description in the  WW Physics section of this
report for what concerns  charged current processes and some general features 
of the program. 

For all processes with $b\ \bar b$ in the final state together with
$\mu \bar \mu$, $\nu_{\mu} \bar \nu_{\mu}$, $\nu_e \bar \nu_e$ and $b \ \bar b$,
finite $b$ masses are properly taken into account both in the phase space and in
matrix elements. Higgs contributions are of course included.

Full tree level matrix elements for these  processes (as well as for
all other four fermion final states) are computed by means of subroutines 
which make use of the helicity formalism of ref.~\cite{method}, which is
particularly suited for treating massive fermion processes.
The code for them has been written semi automatically through the set
of routines PHACT~\cite{phact} ({\bf P}rogram for {\bf H}elicity 
{\bf A}mplitudes                                    
{\bf C}alculations with {\bf T}au matrices)  which implements the
method in a fast and efficient way.
In the above  formalism, eigenstates of the fermion propagators are 
used to simplify   matrix expressions. These eigenstates are chosen to be
generalizations of the spinors used in ref.~\cite{ks}. With the introduction of
so called tau matrices~\cite{method}, the numerators of fermion propagators 
have a very simple expression also in the massive case and one does not have
to care about the various mass terms.  The computation
of fermion lines reduces to evaluating  the matrices corresponding to
insertions of vector or scalar lines and combining them together. 
 The program PHACT writes automatically the optimized 
FORTRAN code necessary for every insertion and every combination, 
given the names of the vectors, couplings, etc. It turns out that the massive 
case is not more complicated than the massless one. Only more 
helicity indices are of course needed. As a consequence, the codes for massive
amplitudes  written in this way are not much slower, as it is normally the
case, than those with massless fermions.
    
The user has the choice among three different ways of sampling the phase space,
in order to take into account 
the peak structure of the  Higgs signal and of the other resonating diagrams 
of the background. The adaptive routine  VEGAS~\cite{vegas} is used for 
integrating over the phase space.                           
        
\noindent {\bf Features of the program.}
WPHACT is a Monte Carlo program. The integration is performed by
VEGAS~\cite{vegas}.  For all phase spaces used, all momenta are explicitly 
computed in terms of the  integration variables. This implies that any cut 
can be implemented, and it can be easily used also as an event generator.
The events obtained in this way are of course weighted. 
Distributions for any variable can also be easily implemented, even if
no automatic implementation of distributions has yet been introduced.

All final states  computed by WPHACT correspond to four fermions. Thus
no stable Z or Higgs  are allowed in the final state. They are always 
considered as virtual particles. The Higgs  decay particles
are always treated as massive, both in the matrix elements
of signal and background and in the phase space. 
All tree level QCD background processes 
($ {\cal O} (\alpha\ \as )$) leading to four-quark final states
are completely taken into account.                      
Initial state QED radiation is included through structure functions 
${\cal O} (\alpha ^2)$.  
Anomalous gauge boson couplings are also present, if required.
FSR is not implemented and no interface to hadronization is available.
                                                            
It is easy to obtain contributions from different set of diagrams, as 
every diagram is evaluated individually for all helicity configuration and then
summed to the others before squaring and summing over helicity configurations.
In particular contributions to Higgs signal, background and their
interference can be evaluated separately. 

We give some indicative values about
the running time on an ALPHA AXP 2100/4 OVMS, in the massive case:

CPU time  per call for $e^+ e^- \rightarrow b \bar{b} b \bar{b}$
     Higgs signal with ISR: $3.0\times 10^{-4}$ sec.

CPU time  per call for $e^+ e^- \rightarrow b \bar{b} b \bar{b}$    
     Higgs background with ISR: $1.3\times 10^{-3}$ sec.

CPU time  per call for $e^+ e^- \rightarrow b \bar{b} \mu^+ \mu^-$    
     Higgs signal with ISR: $9.0\times 10^{-5}$ sec.

CPU time  per call for $e^+ e^- \rightarrow b \bar{b} \mu^+ \mu^-$    
     Higgs background with ISR: $6.3\times 10^{-4}$ sec.
                                
For the same processes without ISR,  CPU time per call is 
about 20\% less. On a VAXstation 4000/90  CPU time for these
programs has to be multiplied  approximately by a factor $5$.

At LEP2 energies, 2.5 M  calls (about 13 minutes for the first process and 
4 minutes for the second one)  are used on ALPHA AXP
to obtain Higgs signal with ISR  cross section with a typical estimated error of 
about  $1 \times 10^{-4}$.  The same processes can be evaluated in about 
1.5 minutes  and 15 sec. respectively  with 0.2 M calls at permill level.
At this level 2.5 M calls (30~minutes) are necessary for   
$e^+ e^- \rightarrow b \bar{b} \mu^+ \mu^-$ Higgs background with 
ISR while 16 M calls (6 hours) are needed for $e^+ e^- \rightarrow b
\bar{b}b \bar{b} $ Higgs background  with ISR.

\noindent {\bf How the code works.}
The  variables which parametrize 
the phase space are: the  masses 
of the two virtual Z's (or those of the virtual Higgs and Z),
the angle of the two  particles with respect to 
the beam, the decay angles in their rest frames, and $x_1$, $x_2$, the 
fractions of momenta carried by the electrons.  Appropriate changes of 
variables to optimize the sampling of the 
peaks in $x_1$, $x_2$, $M_H$ and $M_Z$ lead to 
the actual integration variables. 
For every point chosen by the integration routine, the full set of four 
momenta are reconstructed and passed to the
subroutine which evaluates the differential cross section 
with the helicity amplitude formalism. For every point in the integration 
variables, i.e. for every set of four momenta chosen, VEGAS gives a weight
which can be used together with the value of the cross section for producing
distributions. 

Three different ways of sampling the massive phase space are  available, which
are appropriate for different peaking  structures. We can classify them as
double resonant, single resonant and non resonant. We have verified that
normally the double resonant phase space is accurate enough. The other two can
be used to study contributions of a particular subset of diagrams. It is better
to  run the Higgs signal and background separately, adding the results, as the
change of variables necessary to take care of the resonances of the two
contributions depends on their masses.  The interference is normally added to
the background, but it can be separated and evaluated by itself. 

The $e^+ e^- \rightarrow b \bar b b\bar b $ is in principle a little
more complicated to integrate than processes with only one pair of $b$'s in
the final state. This is due to the presence of identical particles in the final
state which implies that each $b$ can be resonating in some diagrams
with the first $ \bar b$ and in others with the second one. This further 
complicates the subdivision in double resonant contributions, but we have
reduced it to the simpler cases just exploiting the symmetries of the problem.
This simplification is exact only in the symmetric case. One cannot thus
evaluate at present the four $b$ processes with cuts which are not symmetric
under the exchange of the two $b$'s or of the two $\bar b$'s among themselves.
A cut which does not fulfill the above requirement is in any case unphysical. 

After every iteration the integration routine readjusts the grid in the 
space of integration variables, in order to concentrate evaluations of
the integrands in those regions where the integrand is larger in magnitude.
It is advised to use a first iteration  with few points to "thermalize". 
   
\noindent {\bf Input parameters, flags, etc.}
The standard input parameters are $M_H$, $M_W$, $M_Z$, $M_b$, $\alpha$, 
$\as$. In the tuned comparisons presented in sect ~\ref{higgscomp}
$\sin^2 \theta_W$ has been given
as an input,  while it is usually derived from the relation 
$\sin^2 \theta_W=1-M_W^2/M_Z^2$.

The main flag of the program is {\tt ich}, which chooses among different 
final states. 
Other flags allow to compute  with or without ISR ({\tt isr}), to choose
among signal, background and interference ({\tt isig}), and to choose 
whether or not to use some thermalizing iterations ({\tt iterm}). 
The number of                           
iterations ({\tt itmx}) and of  points per iteration ({\tt ncalls}) for
the thermalizing phase as well as for the normal one and the accuracy
required ({\tt acc}) are read from the input.

\noindent {\bf Output.}
The output is just the standard VEGAS output, from which one can read the 
final result and estimated statistical error,  as well as the 
result and error for every iteration. Results with big oscillations among
different iterations and corresponding big reported $\chi^2$ have to be
discarded and simply mean that the number of evaluations per iteration was
not sufficient for the integrand.

\noindent {\bf Concluding remarks.}
As already stated, WPHACT makes use of matrix elements which are suitable
for massive fermion calculations. One may question how big the mass effects
are for Higgs physics. Using WPHACT one can verify that they are normally
at the percent level. They however depend  on the Higgs mass, and 
expecially on the cuts introduced. These may change
the expected dependence, and any set of realistic cuts has to be studied
independently with programs which take masses into account.

WPHACT  does not make use of any library, has proven to be 
reliable over a vast range of statistical           
errors and can compute in short time  exact massive 
processes of interest for Higgs physics at $e^+ e^-$ colliders.
\subsection{WTO}
\begin{center}                             
\begin{tabular}{ll}
Program name:          & WTO \\
Author:                & G. Passarino -- {\tt giampiero@to.infn.it} \\
Availability:          & \\
Documentation:         & \\
\end{tabular}
\end{center}

WTO is a {\it quasi-analytical, deterministic} code for computing
observables related to the process $e^+e^- \to {\bar f_1}f_2{\bar f}_3f_4$. The
full matrix elements are used and in the present version the 
following final states are accessible (see~\cite{class} for a general
classification):
\begin{enumerate}
\item CC3, CC11, CC20 
\item NC19, NC24, NC32
\item NC21 (= NC19 + Higgs), NC25 (= NC24 + Higgs)
\item MIX43
\end{enumerate}
Further extensions will be gradually implemented.
To fully specify WTO's setup an option must be chosen for the renormalization 
scheme (RS). One has:
\begin{enumerate}
\item the option commonly used for tuned comparisons, i.e.
\begin{equation}
s_{_W}^2 = {{\pi\alpha(2\wm)}\over {{\sqrt 2}\gf\wm^2}},  \qquad
g^2 = {{4\pi\alpha(2\wm)}\over {s_{_W}^2}}
\end{equation}
\item or the default,
\begin{equation}
s_{_W}^2 = 1 - {{\wm^2}\over {\zm^2}}, \qquad
g^2 = 4{\sqrt 2}\gf\wm^2
\end{equation}
\end{enumerate}
where $\alpha^{-1}(2\wm) = 128.07$ and $\gf$ is the Fermi coupling constant.
Final state QCD corrections are not taken into account in the present
version, but for the Higgs width.
A more complete description of WTO is given in ref.~\cite{wwsection}. 
                                                                    
Among all four-fermion processes included in WTO~\cite{wwsection}, 
those of relevance to Higgs physics are:                        
\begin{equation}
e^+e^- \to {\bar b}b{\bar X}X,  \qquad \left(X = l, q\not= b\right)
\end{equation}
The matrix elements are obtained with the helicity method described
in~\cite{cpb}.  The whole answer is written in terms of invariants, i.e.                           
\begin{eqnarray}                                                                                   
e^+(p_+)e^-(p_-) &\to& f(q_1){\bar f}(q_2)f'(q_3){\bar f}'(q_4),  \\
x_{ij}s &=& -\left(q_{i-2} + q_{j-2}\right)^2,  \quad
x_{1i}s = -\left(p_+ + q_{i-2}\right)^2,  \\
x_{2i}s &=& -\left(p_- + q_{i-2}\right)^2,  \quad
s_1s^2 = \epsilon\left(p_+,p_-,q_1,q_2\right), \dots
\end{eqnarray}
and the integration variables are chosen to be
\begin{eqnarray}
m_-^2 &=& x_{24}, \quad m_+^2 = x_{56}, \quad
M_0^2 = x_{45},  \quad m_0^2 = x_{36}, \\
m^2 &=& x_{35}, \quad t_1 = x_{13}, \quad t_{_W} = x_{13} + x_{14}
\end{eqnarray}
The convention for the final states in WTO is: $e^+e^- \to 1+2+3+4$.
For CC processes $1=d, 2={\bar u}, 3=u', 4={\bar d}'$, with
$u = \nu,u,c$ and $d = l,d,s,b$. For NC processes the adopted
convention is $1=f, 2={\bar f}, 3=f'$ and $4={\bar f}'$.
Initial state QED radiation is included through the Structure Function approach 
up to $O(\alpha^2)$. The code will return results according to three
(pre-selected) options, i.e $\beta^2\eta$ (default)~\cite{topv}, 
$\beta^3$~\cite{cpcww} and $\beta\eta^2$~\cite{exca} where
\begin{equation}
\beta = 2\,\frac{\alpha}{\pi}\,\left(\log\frac{s}{m_e^2} - 1\right),
\qquad \eta = 2\,\frac{\alpha}{\pi}\,\log\frac{s}{m_e^2} 
\end{equation}
When initial state QED radiation is included there are two additional 
integrations over the fractions of the beam energies lost through radiation,
$x_{\pm}$. 
This description of the phase space gives full cuts-availability through
an analytical control of the boundaries of the phase space. Upon
specification of the input flags it is therefore possible to cut on
all final state invariant masses, all (LAB) final state energies $E_i, i=1,4$,
all (LAB) scattering angles, $\theta_i, i=1,4$, all (LAB) final state angles, 
$\psi_{ij}, i,j=1,4$.
Both the matrix elements and the phase space are given for massless fermions.
There is no interface with hadronization.
The integration is performed with the help of the NAG routine 
D01GCF. This routine uses the Korobov-Conroy number theoretic approach with a 
MC error estimate arising from converting the number theoretic formula for the 
$n$-cube $[0,1]^n$ into a stochastic integration rule. This allows a `standard 
error' to be estimated. Prior to a call to D01GCF the peak structure of the 
integrand is treated with the appropriate mappings.
The typical process considered belong to the NC21 or NC25 classes. In WTO both 
the phase space and the matrix elements are written for massless fermions, thus
there is no interference between the Higgs signal and the background, making 
particularly easy to include the Higgs boson.
The pole quark masses are specified in a DATA BLOCK as $m_q(m_q^2)$ and the
code will convert them internally into running masses, i.e. $m_q(m_H^2)$.
Whenever needed the input parameter $\alpha_s(\wm)$ is also converted into
$\alpha_s(m_H^2)$. The obtained $m_q(m_q^2)$ are then used to generate the 
couplings $H \to {\bar q}q$. The Higgs width is computed as
$\Gamma_H = \Gamma(H \to \tau^+\tau^-, {\bar c}c, {\bar b}b, gg)$ and upon
proper initialization of the corresponding flag final state QCD corrections are
applied.

Numerical input parameters such as $\alpha(0), \gf, \zm, \wm, \dots$ are
stored in a BLOCK DATA.
There are various flags to be initialized to run WTO. Here follows a short 
description of the most important ones:                     
\begin{itemize}
\item[\verb+NPTS  +] - INTEGER, NPTS=1,10 chooses the actual number of points for 
applying the Korobov-Conroy number theoretic formulas. The built-in choices
correspond to to a number of actual points ranging from 2129 up to 5,931,551.
\item[\verb+NRAND +] - INTEGER, NRAND specifies the number of random samples to be 
generated in the error estimation (usually $5-6$).
\item[\verb+OXCM  +] - CHARACTER*1, the main decision branch for the process: [C(N)] for 
CC,(NC)~\cite{class}.
\item[\verb+OTYPEM+] - CHARACTER*4,Specifies the process, i.e. CC3, CC11, CC20 for 
CC processes and NC19, NC24, NC21, NC25, NC32 for NC processes.
\item[\verb+IOS   +] - INTEGER, two options [$1,2$] ($1=$default for tuned comparisons) 
for the RS.
\item[\verb+IOSF  +] - INTEGER, three options [$1-3$] for the $\eta-\beta$ choice in the 
structure functions.
\item[CHDM$\dots$] - REAL, Electric charges, third component of isospin for 
the final states.
\end{itemize}
WTO is a robust one call - one result code, thus in the output one gets a list
of all relevant input parameters plus the result of the requested
observable with an estimate of the numerical error. A very 
rough estimate of the theoretical error (very subjective to say the least)
can be obtained by repeating runs with different IOS, IOSF options.
After the following initialization:
\begin{verbatim}                  

7 6                     !  NPTS NRAND
175.d0                  !  E_CM OF PROCESS
n                       !  NC PROCESS
nc25                    !  CLASS = NC25
l                       !  MU
65.d0                   !  M_H(GEV)
0.12d0                  !  ALPHA_S(M_W)
y                       !  FS QCD
y                       !  H --> GG INCLUDED
1 1                     !  IOS IOSF
hc                      !  BUILT-IN CHOICE OF CUTS 
hl                      !                    "
-1.d0 -0.33333333333d0  !  CHARGES: F=MU, FP=B
-0.5d0 -0.5d0           !  ISOSPINS
1.d0 3.d0               !  COLOR FACTORS

\end{verbatim}
corresponding to the process for $e^+e^- \to \mu^+\mu^- {\bar b}b$ with 
$\zm - 25\,$GeV$ < M_{\nu\nu} < \zm + 25\,$ GeV, 
$M_{{\bar b}b} > 50\,$ GeV, the typical output will look as follows:
\begin{verbatim}

This run is with: 

NPTS         =  7
NRAND        =  6

E_cm (GeV) =          0.17500E+03
beta       =          0.11376E+00 sin^2     =          0.23103E+00
M_W  (GeV) =          0.80230E+02 M_Z (GeV) =          0.91189E+02
G_W  (GeV) =          0.20337E+01 G_Z (GeV) =          0.24974E+01
M_H  (GeV) =          0.65000E+02 G_H (MeV) =          0.15865E+01

m_b(M_H) (GeV) =          0.29168E+01
m_c(M_H) (GeV) =          0.64862E+00
alpha_s(M_H)   =          0.12402E+00

nc25-diagrams : charges    -1.0000   -0.3333
                isospin    -0.5000   -0.5000

On exit IFAIL = 0 - Cross-Section 

CPU time  28 min  37 sec, sec per call  =  0.286E-02
# of calls      =     599946

(Signal) sigma  =  0.2766804E-01   +-          0.1188170E-04

Rel. error of      0.043 %
\end{verbatim}
\subsection{Comparisons among the programs}
\label{higgscomp}
In this section we present some ``tuned'' comparisons between
semianalytical/deterministic and Monte Carlo codes for Higgs searches.
In the case of SM Higgs production, we will consider the following processes:
\br
e^+e^- &\to &b \bar b    \mu^+\mu^- \nonumber \\
e^+e^- &\to &b \bar b    \nu_{\mu} \bar\nu_{\mu} \nonumber \\
e^+e^- &\to &b \bar b    \nu_{e} \bar\nu_{e}. \nonumber \\            
\er                                          
The selection criteria adopted involve only invariant-mass cuts, in
order to allow also some semianalytical approaches to appear in the
comparisons. These cuts are:
$M_Z - 25$~GeV $\leq$ $m_{l \bar l}$ $\leq$ $M_Z + 25$~GeV;
$m_{b \bar b} \geq $  50 GeV. 
Cross section values for different beam energies and different Higgs masses are
given in Tables~\ref{htab_1}~--~\ref{htab_12}. As a reference, the last column
of each Table contains the cross sections 
in absence of Higgs signal (pure          
4-fermion background).
The results of the pure non-Higgs channels obtained by the EXCALIBUR
\cite{wwsection}  and FERMISV\footnote{The numbers for FERMISV were kindly
generated by P. Janot.}~\cite{fermisv} code are also shown.                      
The input parameters used in these Tables are the {\sl STANDARD LEP2        
INPUT} \cite{wwsection}. The only                            
exception is the choice of fermion masses. Since the 
$H \to f \bar f$ coupling constant is proportional to $m_f$,
the choice adopted here \cite{higgssection} is to use       
running fermion masses $m_f = m_f(Q^2 = m_H^2)$ in the Higgs-boson
coupling~\cite{higgssection}. 
The codes which can evaluate massive amplitudes (CompHEP, GENTLE/4fan
and WPHACT), adopt however different prescriptions for the choice of
the $b$ mass appearing in the phase space and in the matrix         
elements. For example, WPHACT can fix this to be the pole mass, while
GENTLE/4fan adopts the same value used for the coupling to the Higgs.
The suffix added to the results of the CompHEP and WPHACT programs refers to
the value of the $b$ quark mass used in the evaluation of the {\em production}
matrix elements.
The effect of the complete inclusion of $b$-masses in the matrix
elements is clearly visible from the Tables, although it never exceeds the 
\%~level.
                                             
Few comments on the results are in order.
With the exception of HZHA, which does not 
include the full set of SM background diagrams, the agreement between the Higgs
codes presented in the Tables is systematically at the level of 1\% or better. 
The exceptions are the processes with $\nu_e \bar\nu_e$ in the final state,
where CompHEP differs by approximately 2\% from the other codes (see
Table~\ref{htab_5} and \ref{htab_6}). Notice that
this is the channel where the difference between having and not having the full
set of SM diagrams is potentially the largest, as indicated by the results of
HZHA, which can differ from the other codes by up to 20\%.
While discrepancies at the \%~level are of the order of the net uncertainty
coming from higher order corrections, it is clear that they should be studied
further in order to make any future full NLO result meaningful.
At the same time, it is important to point out that the impact of the
discrepancies we found on the discovery potential of LEP2 is minimal. Whether
these differences could affect the extraction of Higgs properties after its
discovery at LEP2 is an interesting question, which however will require
further work to be answered.
                            
{\renewcommand{\arraystretch}{1.3}
\begin{table}[hbtp]              
\footnotesize
\begin{center}
\begin{tabular}{|l|l|l|l|l|}
\hline 
$m_H$~(GeV)    & 65          & 90          & 115         & $\infty$  \\
\hline
CompHEP$_0$    & 32.487(63)  & 1.593(03)   & 1.059(02)   & 1.059(02) \\
CompHEP$_{4.7}$& 32.474(63)  & 1.578(03)   & 1.046(02)   & 1.046(02) \\
EXCALIBUR      & ---         & ---         & ---         & 1.0594(03)\\
FERMISV        & ---         & ---         & ---         & 0.931(22) \\
GENTLE/4fan    & 32.7148(33) & 1.59930(16) & 1.05949(11) & 1.05944(11)\\
HIGGSPV        & 32.714(27)  & 1.607(08)   & 1.060(02)   & 1.049(07) \\
HZHA           & 32.435(33)  & 1.570(33)   & 1.056(33)   & 1.056(33) \\
WPHACT$_{4.7}$ & 32.5604(66) & 1.58552(62) & 1.04684(56) & 1.04679(55)\\ 
WPHACT$_0$     & 32.7141(68) & 1.59946(64) & 1.05953(56) & 1.05948(56)\\
WTO            & 32.7268(51) & 1.5980(13)  & 1.0582(12)  & 1.0581(12)\\
\hline
\end{tabular}
\end{center}
\ccaption{}{\label{htab_1}
$\sigma(e^+ e^- \to \mu^+ \mu^- b \bar b )$ (fb) at
$ E_{cm} = 175$~GeV. No ISR.}

\end{table}
}

{\renewcommand{\arraystretch}{1.3}
\begin{table}[hbtp]              
\footnotesize
\begin{center}
\begin{tabular}{|l|l|l|l|l|}
\hline
$m_H$~(GeV)    & 65            & 90           & 115         &  $\infty$ \\
\hline
CompHEP$_0$    & 37.264(58)    & 24.395(46)   & 10.696(13)  & 10.634(13) \\
CompHEP$_{4.7}$& 37.147(58)    & 24.279(46)   & 10.580(13)  & 10.518(13) \\
EXCALIBUR      & ---           & ---          & ---         & 10.6398(15)\\
FERMISV        & ---           & ---          & ---         & 9.49(23) \\
GENTLE/4fan    & 37.3975(37)   & 24.4727(25)  & 10.7022(11) & 10.6401(11)\\
HIGGSPV        & 37.393(27)    & 24.490(21)   & 10.694(16)  & 10.65(05) \\
HZHA           & 36.79(13)     & 23.53(13)    & 10.28(13)   & 10.22(13) \\
WPHACT$_{4.7}$ & 37.1634(64)   & 24.3245(40)  & 10.5863(24) & 10.5243(24)\\
WPHACT$_0$     & 37.3990(64)   & 24.4727(40)  & 10.7027(24) & 10.6407(24)\\
WTO            & 37.4099(32)   & 24.4765(42)  & 10.7036(21) & 10.6416(21)\\
\hline
\end{tabular}
\end{center}
\ccaption{}{\label{htab_2}
$\sigma(e^+ e^- \to \mu^+ \mu^- b \bar b )$ (fb) at
$ E_{cm} = 192$~GeV. No ISR.}
\end{table}
}
{\renewcommand{\arraystretch}{1.3}
\begin{table}[hbtp]              
\footnotesize
\begin{center}
\begin{tabular}{|l|l|l|l|l|}
\hline
$m_H$~(GeV)      & 65           & 90          & 115           &  $\infty$ \\
\hline
CompHEP$_0$      & 64.14(15)    & 2.341(07)   & 1.279(04)     & 1.279(04) \\
CompHEP$_{4.7}$  & 64.12(15)    & 2.325(07)   & 1.263(04)     & 1.263(04) \\
EXCALIBUR        & ---          & ---         & ---           & 1.2916(04)\\
FERMISV          & ---          & ---         & ---           & 1.195(26) \\
GENTLE/4fan      & 64.2407(64)  & 2.36582(24) & 1.29239(13)   & 1.29229(13)\\
HIGGSPV          & 64.199(60)   & 2.375(19)   & 1.293(09)     & 1.286(14) \\
HZHA             & 63.99(02)    & 2.258(18)   & 1.230(18)     & 1.230(18) \\
WPHACT$_{4.7}$   & 63.941(14)   & 2.3473(10)  & 1.27611(80)   & 1.27601(80)\\
WPHACT$_0$       & 64.238(14)   & 2.3661(10)  & 1.29237(82)   & 1.29227(82)\\
WTO              & 64.262(11)   & 2.36583(93) & 1.29210(92)   & 1.2950(20)\\
\hline
\end{tabular}
\end{center}
\ccaption{}{\label{htab_3}
$\sigma(e^+ e^- \to \nu_{\mu} {\bar \nu_{\mu}} b \bar b )$ (fb) at
$ E_{cm} = 175$~GeV. No ISR.}
\end{table}
}

{\renewcommand{\arraystretch}{1.3}
\begin{table}[hbtp]              
\footnotesize
\begin{center}
\begin{tabular}{|l|l|l|l|l|}
\hline
$m_H$~(GeV)     & 65           & 90           & 115          &  $\infty$ \\
\hline
CompHEP$_0$     & 72.64(19)    & 47.02(14)    & 19.76(08)    & 19.62(07) \\
CompHEP$_{4.7}$ & 72.41(19)    & 46.79(14)    & 19.53(08)    & 19.41(07) \\
EXCALIBUR       & ---          & ---          & ---          & 19.7131(40)\\
FERMISV         & ---          & ---          & ---          & 18.57(62) \\
GENTLE/4fan     & 72.9256(73)  & 47.2239(47)  & 19.8405(20)  & 19.7171(20)\\
HIGGSPV         & 72.867(63)   & 47.225(50)   & 19.786(42)   & 19.67(06) \\
HZHA            & 72.83(21)    & 46.31(21)    & 19.82(21)    & 19.71(21) \\
WPHACT$_{4.7}$  & 72.475(16)   & 46.944(12)   & 19.625(11)   & 19.502(10)\\
WPHACT$_0$      & 72.927(16)   & 47.222(12)   & 19.841(11)   & 19.717(11)\\
WTO             & 72.961(11)   & 47.2341(40)  & 19.8394(14)  & 19.7200(70)\\
\hline
\end{tabular}
\end{center}
\ccaption{}{\label{htab_4}
$\sigma(e^+ e^- \to \nu_{\mu} {\bar \nu_{\mu}} b \bar b )$ (fb) at
$ E_{cm} = 192$~GeV. No ISR.}
\end{table}
}
{\renewcommand{\arraystretch}{1.3}
\begin{table}[hbtp]              
\footnotesize
\begin{center}
\begin{tabular}{|l|l|l|l|l|}
\hline
$m_H$~(GeV)     & 65             & 90            & 115        &  $\infty$ \\
\hline
CompHEP$_0$     & 70.26(20)      & 5.03(02)      & 1.073(04)  & 1.073(04) \\
CompHEP$_{4.7}$ & 70.24(20)      & 5.02(02)      & 1.059(04)  & 1.059(04)  \\
EXCALIBUR       & ---            & ---           & ---        & 1.0796(03)\\
FERMISV         & ---            & ---           & ---        & 1.195(26) \\
HIGGSPV         & 71.727(34)     & 5.100(05)     & 1.081(01)  & 1.077(06) \\
HZHA            & 69.98(18)      & 3.572(18)     & 1.230(18)  & 1.230(18) \\
WPHACT$_{4.7}$  & 71.366(26)     & 5.0762(22)    & 1.06615(87)& 1.06602(87)\\
WPHACT$_0$      & 71.694(27)     & 5.0996(23)    & 1.08027(89)& 1.08013(89)\\
WTO             & 71.679(14)     & 5.0997(15)    & 1.07978(81)& 1.0820(20)\\
\hline
\end{tabular}
\end{center}
\ccaption{}{\label{htab_5}
$\sigma(e^+ e^- \to \nu_e {\bar \nu_e} b \bar b )$ (fb) at
$ E_{cm} = 175$~GeV. No ISR.}
\end{table}
}
{\renewcommand{\arraystretch}{1.3}
\begin{table}[hbtp]              
\footnotesize
\begin{center}
\begin{tabular}{|l|l|l|l|l|}
\hline
$m_H$~(GeV)     & 65             & 90            & 115           &  $\infty$ \\
\hline
CompHEP$_0$     & 79.01(24)      & 52.37(18)     & 20.82(08)     & 19.89(07) \\
CompHEP$_{4.7}$ & 78.79(24)      & 52.15(18)     & 20.60(08)     & 19.67(07)  \\
EXCALIBUR       & ---            & ---           & ---           & 19.9463(44)\\
FERMISV         & ---            & ---           & ---           & 18.57(62) \\
HIGGSPV         & 80.628(32)     & 53.353(21)    & 20.907(13)    & 19.95(10) \\
HZHA            & 80.99(21)      & 49.80(21)     & 20.26(21)     & 19.71(21) \\
WPHACT$_{4.7}$  & 80.122(34)     & 53.039(19)    & 20.673(12)    & 19.736(12)\\
WPHACT$_0$      & 80.611(34)     & 53.335(19)    & 20.893(12)    & 19.955(10)\\
WTO             & 80.629(32)     & 53.3468(63)   & 20.8883(15)   & 19.9540(50)\\
\hline
\end{tabular}
\end{center}
\ccaption{}{\label{htab_6}
$\sigma(e^+ e^- \to \nu_e {\bar \nu_e} b \bar b )$ (fb) at
$ E_{cm} = 192$~GeV. No ISR.}
\end{table}
}
{\renewcommand{\arraystretch}{1.3}
\begin{table}[hbtp]              
\footnotesize
\begin{center}
\begin{tabular}{|l|l|l|l|l|}
\hline
$m_H$~(GeV)    & 65            & 90          & 115          &  $\infty$ \\
\hline
EXCALIBUR & --- & --- & --- & 0.8256(04) \\
FERMISV & ---  & --- & --- & 0.745(19) \\
GENTLE/4fan    & 28.4273(28)   & 1.22507(12) & 0.824890(82) & 0.824849(82)\\
HIGGSPV        & 28.437(14)    & 1.224(02)   & 0.8248(06)   & 0.817(06) \\
HZHA           & 28.317(27)    & 1.252(27)   & 0.860(27)    & 0.860(27) \\
WPHACT$_{4.7}$ & 28.305(17)    & 1.21406(85) & 0.81492(81)  & 0.81489(81)\\
WPHACT$_0$     & 28.437(17)    & 1.22479(70) & 0.82472(65)  & 0.82468(65)\\
WTO            & 28.456(12)    & 1.2241(16)  & 0.8232(15)   & 0.8232(15)\\
\hline
\end{tabular}
\end{center}
\ccaption{}{\label{htab_7}
$\sigma(e^+ e^- \to \mu^+ \mu^- b \bar b )$ (fb) at
$ E_{cm} = 175$~GeV. ISR included.}
\end{table}
}
{\renewcommand{\arraystretch}{1.3}
\begin{table}[hbtp]              
\footnotesize
\begin{center}
\begin{tabular}{|l|l|l|l|l|}
\hline
$m_H$~(GeV)    & 65            & 90           & 115          &  $\infty$ \\
\hline
EXCALIBUR & --- & --- & --- & 8.4306(29) \\
FERMISV & ---  & --- & --- & 7.90(27) \\
GENTLE/4fan    & 33.7575(34)   & 19.4717(19)  & 8.47729(85)  & 8.43290(84)\\
HIGGSPV        & 33.759(12)    & 19.480(09)   & 8.483(05)    & 8.44(05) \\
HZHA           & 33.48(11)     & 18.91(11)    & 8.31(11)     & 8.27(11) \\
WPHACT$_{4.7}$ & 33.547(15)    & 19.3515(90)  & 8.3842(56)   & 8.3400(56)\\
WPHACT$_0$     & 33.752(16)    & 19.4692(91)  & 8.4767(57)   & 8.4324(57)\\
WTO            & 33.777(10)    & 19.4856(83)  & 8.4851(78)   & 8.4409(78)\\
\hline
\end{tabular}
\end{center}
\ccaption{}{\label{htab_8}
$\sigma(e^+ e^- \to \mu^+ \mu^- b \bar b )$ (fb) at
$ E_{cm} = 192$~GeV. ISR included.}
\end{table}
}
{\renewcommand{\arraystretch}{1.3}
\begin{table}[hbtp]              
\footnotesize
\begin{center}
\begin{tabular}{|l|l|l|l|l|}
\hline
$m_H$~(GeV)    & 65             & 90            & 115           &  $\infty$ \\
\hline
EXCALIBUR & --- & --- & --- & 0.9900(05) \\
FERMISV & ---  & --- & --- & 0.928(23) \\
GENTLE/4fan    & 55.9190(56)    & 1.78649(18)   & 0.990681(99)  & 0.990600(10)\\
HIGGSPV        & 55.899(29)     & 1.786(05)     & 0.991(02)     & 0.991(12) \\
HZHA           & 55.863(14)     & 1.733(14)     & 0.949(14)     & 0.949(14) \\
WPHACT$_{4.7}$ & 55.644(34)     & 1.77146(97)   & 0.97777(83)   & 0.97770(83)\\
WPHACT$_0$     & 55.901(34)     & 1.7858(10)    & 0.99028(84)   & 0.99021(84)\\
WTO            & 55.947(27)     & 1.7857(14)    & 0.9894(13)    & 0.9893(13)\\
\hline
\end{tabular}
\end{center}
\ccaption{}{\label{htab_9}
$\sigma(e^+ e^- \to \nu_{\mu} {\bar \nu_{\mu}} b \bar b )$ (fb) at
$ E_{cm} = 175$~GeV. ISR included.}
\end{table}
}
{\renewcommand{\arraystretch}{1.3}
\begin{table}[hbtp]              
\footnotesize
\begin{center}
\begin{tabular}{|l|l|l|l|l|}
\hline
$m_H$~(GeV)& 65 & 90 & 115 &  $\infty$ \\
\hline
EXCALIBUR & --- & --- & --- & 15.5420(64) \\
FERMISV & ---  & --- & --- & 15.14(56) \\
GENTLE/4fan    & 65.9061(66)    & 37.4957(37)    & 15.6302(16)  & 15.5421(16)\\
HIGGSPV        & 65.895(27)     & 37.504(20)     & 15.629(13)   & 15.51(06) \\
HZHA           & 65.60(14)      & 36.45(14)      & 15.25(14)    & 15.17(14) \\
WPHACT$_{4.7}$ & 65.500(31)     & 37.270(18)     & 15.460(12)   & 15.372(12)\\
WPHACT$_0$     & 65.894(31)     & 37.491(18)     & 15.631(12)   & 15.543(12)\\
WTO            & 65.922(27)     & 37.5201(96)    & 15.6356(50)  & 15.5474(50)\\
\hline
\end{tabular}
\end{center}
\ccaption{}{\label{htab_10}
$\sigma(e^+ e^- \to \nu_{\mu} {\bar \nu_{\mu}} b \bar b )$ (fb) at
$ E_{cm} = 192$~GeV. ISR included.}
\end{table}
}
{\renewcommand{\arraystretch}{1.3}
\begin{table}[hbtp]              
\footnotesize
\begin{center}
\begin{tabular}{|l|l|l|l|l|}
\hline
$m_H$~(GeV)& 65 & 90 & 115 &  $\infty$ \\
\hline
EXCALIBUR & --- & --- & --- & 0.8382(05) \\
FERMISV & ---  & --- & --- & 0.928(23) \\
HIGGSPV        & 62.917(35)     & 3.903(04)    & 0.8398(04)    & 0.844(05) \\
HZHA           & 60.96(14)      & 2.753(14)    & 0.949(14)     & 0.949(14) \\
WPHACT$_{4.7}$ & 62.589(32)     & 3.8858(25)   & 0.82761(65)   & 0.82751(65)\\
WPHACT$_0$     & 62.876(32)     & 3.9037(25)   & 0.83849(66)   & 0.83838(66)\\
WTO            & 62.905(65)     & 3.9056(40)   & 0.8381(13)    & 0.8379(13)\\
\hline
\end{tabular}
\end{center}
\ccaption{}{\label{htab_11}
$\sigma(e^+ e^- \to \nu_e {\bar \nu_e} b \bar b )$ (fb) at
$ E_{cm} = 175$~GeV. ISR included.}
\end{table}
}
{\renewcommand{\arraystretch}{1.3}
\begin{table}[hbtp]              
\footnotesize
\begin{center}
\begin{tabular}{|l|l|l|l|l|}
\hline
$m_H$~(GeV)& 65 & 90 & 115 &  $\infty$ \\
\hline
EXCALIBUR & --- & --- & --- & 15.5974(69) \\
FERMISV & ---  & --- & --- & 15.14(56) \\
HIGGSPV        & 73.051(34)    & 42.682(21)    & 16.275(12)    & 15.78(09) \\
HZHA           & 72.85(14)     & 39.35(14)     & 15.56(14)     & 15.17(14) \\
WPHACT$_{4.7}$ & 72.595(39)    & 42.439(20)    & 16.095(13)    & 15.418(13)\\
WPHACT$_0$     & 73.022(39)    & 42.673(20)    & 16.268(13)    & 15.590(13)\\
WTO            & 73.003(44)    & 42.701(17)    & 16.2675(58)   & 15.5897(58)\\
\hline
\end{tabular}
\end{center}
\ccaption{}{\label{htab_12}
$\sigma(e^+ e^- \to \nu_e {\bar \nu_e} b \bar b )$ (fb) at
$ E_{cm} = 192$~GeV. ISR included.}
\end{table}
}

Only one of the codes presented here (HZHA) allows the generation of SUSY Higgs
bosons. We present a set of cross sections for the                   
$e^+e^- \to b \bar b  b \bar b$ final state for the four cases relative to
the following choice of       
parameters~\cite{higgssection}: 

\begin{tabular}{lll}
(1) & $m_A$ =  75 GeV , & $\tan \beta$ = 30; \\   
(2) & $m_A$ = 400 GeV , & $\tan \beta$ = 30; \\
(3) & $m_A$ =  75 GeV , & $\tan \beta$ = 1.75; \\
(4) & $m_A$ = 400 GeV , & $\tan \beta$ = 1.75. \\
\end{tabular}

The SM input parameters are the same as for the previous comparisons, and
all the $b \bar b$ pairs are required to have $m_{b \bar b} \geq$~20~GeV.
The results are shown in Tables~\ref{htab_13}~--~\ref{htab_16}.
The only comparison possible between the results of HZHA and those of other
codes is for the SM backgrounds.  For these we present, when available,
the separate contribution coming from the                    
purely EW diagrams.
The ${\cal O}(\as\alpha)$ QCD background processes, induced
by gluon splitting diagrams, 
have been evaluated using the exact tree level matrix elements in the case
of the EXCALIBUR and WPHACT. HZHA can evaluate these processes 
only in the parton shower
approximation. Since this approach gives a very low generation efficiency,
the results have a large statistical error. Although consistent with the
exact tree level results, the EW+QCD results from HZHA have therefore not been
included in the Tables.
{\renewcommand{\arraystretch}{1.3}
\begin{table}[hbtp]              
\footnotesize
\begin{center}
\begin{tabular}{|l|l|l|l|l|l|l|}
\hline                         
 & (1) & (2) & (3) &  (4) &  EW & EW+QCD \\
\hline                                   
EXCALIBUR &       ---    &  ---      &  ---       &  ---    &  ---    & 6.859(04) \\
HZHA      & 90.71(46) & 2.902(19) & 158.09(79) & 4.632(54) & 2.760(17) & ---    \\
WPHACT$_0$ &    ---    &  ---      &  ---       &  ---       & 2.580(2) & 6.8589(87)\\
WPHACT$_{4.7}$ &    ---    &  ---      &  ---       &  ---   & ---    & 7.1764(84)\\
\hline                                                                
\end{tabular}
\end{center}
\ccaption{}{\label{htab_13}
$\sigma(e^+ e^- \to  b \bar b b \bar b )$ (fb) at
$ E_{cm} = 175$~GeV. No ISR. See the text for the meaning of the labels (1) -- 
(4). The last two columns refer to the SM background results, separated in pure
EW and full EW+QCD processes.}
\end{table}
}
{\renewcommand{\arraystretch}{1.3}
\begin{table}[hbtp]              
\footnotesize
\begin{center}
\begin{tabular}{|l|l|l|l|l|l|l|}
\hline                         
 & (1) & (2) & (3) &  (4) &  EW & EW+QCD \\
\hline
EXCALIBUR &       ---    &  ---      &  ---       &  ---  &    ---    &  25.933(10) \\
HZHA & 135.17(61) & 23.286(58) & 163.36(75) & 74.04(31)  & 22.816(50) & ---    \\
WPHACT$_0$ &    ---    &  ---      &  ---       &  ---       & 21.897(16)&     
                                                               25.916(18)\\    
WPHACT$_{4.7}$ &    ---    &  ---      &  ---       &  ---   & ---    & 25.946(23)\\
\hline                                                           
\end{tabular}
\end{center}
\ccaption{}{\label{htab_14}
$\sigma(e^+ e^- \to  b \bar b b \bar b )$ (fb) at
$ E_{cm} = 192$~GeV. No ISR. See comments in the previous figure caption.}
\end{table}
}
{\renewcommand{\arraystretch}{1.3}
\begin{table}[hbtp]              
\footnotesize
\begin{center}
\begin{tabular}{|l|l|l|l|l|l|l|}
\hline                         
 & (1) & (2) & (3) &  (4) &  EW & EW+QCD \\
\hline
EXCALIBUR &       ---    &  ---      &  ---       &  --- &  ---    & 8.490(20) \\
HZHA & 76.74(39) & 2.513(20) & 140.20(71) & 3.903(48) & 2.397(18)& ---    \\
WPHACT$_0$ &    ---    &  ---      &  ---       &  ---       & 2.239(2)&  
                                                              8.447(22)\\
WPHACT$_{4.7}$ &    ---    &  ---      &  ---       &  ---   & ---    & 8.993(21)\\
\hline                                                                
\end{tabular}
\end{center}
\ccaption{}{\label{htab_15}
$\sigma(e^+ e^- \to  b \bar b b \bar b )$ (fb) at
$ E_{cm} = 175$~GeV. ISR included. See previous figure caption for comments.}
\end{table}
}%
{\renewcommand{\arraystretch}{1.3}
\begin{table}[hbtp]              
\footnotesize
\begin{center}
\begin{tabular}{|l|l|l|l|l|l|l|}
\hline                         
 & (1) & (2) & (3) &  (4) &  EW & EW+QCD \\
\hline
EXCALIBUR &       ---    &  ---      &  ---       &  ---     &  ---    & 23.045(23) \\
HZHA & 118.60(58) & 18.761(87) & 151.75(75) & 57.74(28)  & 18.384(80) & --- \\
WPHACT$_0$ &    ---    &  ---      &  ---       &  ---  & 17.482(14) & 22.991(34)\\
WPHACT$_{4.7}$ &    ---    &  ---      &  ---       &  ---   & --- & 23.258(37)\\
\hline                                                             
\end{tabular}
\end{center}
\ccaption{}{\label{htab_16}
$\sigma(e^+ e^- \to  b \bar b b \bar b )$ (fb) at
$ E_{cm} = 192$~GeV. ISR included. See previous figure caption for comments.}
\end{table}
}

%
\section{Supersymmetry}                                                    
Supersymmetry \cite{susyrev} is considered to be the most likely candidate for
new physics within the reach of LEP2 \cite{npreport}. 
We assume here that the reader is familiar with the basics of SUSY and with its
most common parameters, and we refer to the review articles in
ref.~\cite{susyrev} or to the New Physics report~\cite{npreport} for definitions
and details. A large body of work has been devoted in the past 10 years to the
development of event generators for the simulation of SUSY signals. Due to the
large interest in the subject, the number of computer programs which calculate
cross sections or generate events is very large; however most of these codes
have  not been designed for distribution, and are not documented here. We will
limit ourselves to present codes which have either been developed during the
Workshop, or which have been discussed and used within the activity of the New
Physics Working group. All of these codes are either already public, or will
soon become. 
             
The main difference between SUSY generators for LEP1 and for LEP2 is related to
the significant r\^ole played at LEP2 by $t$-channel exchange diagrams, which
are almost totally negligible at the $Z$ peak. As a typical example, consider
the chargino pair production. This can proceed via $s$-channel $\gamma$-$Z$
production, or via $t$-channel exchange of the electron scalar-neutrino
($\snu_e$). The interference is always destructive, and can significantly
reduce the production cross section if the sneutrino mass is in the 50--100 GeV
region. Another example, documented in the New Physics section of this report
\cite{npreport}, is that of the scalar electron production, where the
$t$-channel exchange of a neutralino can either decrease or increase the rates.

Alhtough documented only in part in this report, extensive cross checks among
the different codes used by the experimental groups have been performed. These
checks included the study of the proper inclusion of $t$-channel diagrams,
of the dependence of cross sections on the 
parameters of the models, as well as studies of kinematical distributions and
of the effects of the initial state radiation (ISR). Comparisons of decay
branching ratios(BR) for unstable particles have also been performed. All tests
have been pursued until agreement at the percent level was achieved.

In most SUSY generators, the emphasis is placed on covering
as many processes as possible in a unified framework. By doing so, the simplest
approaches have often been pursued. For example, it is generally assumed that
production and decay of SUSY states can be factorized, therefore neglecting
possible initial--final state spin correlations. This choice is forced upon us
by the multitude of possible decays which each SUSY particle has allowed as
soon as the parameters of the theory are slightly changed. Each decay channel
would in principle call for a new evaluation of matrix elements with many-body
final states, including the interference with SM processes and possibly with
other SUSY channels. The multitude of channels to be considered for a generic
point in parameter space is such that a thorough evaluation of the full matrix
elements for all SUSY particles has never been carried out, and finds no place
in any multi-purpose SUSY event generator. In order to assess the limit of this
approach, several groups have started working on more specific channels, where
the structure of the final state is better determined and where full
calculations can be performed and compared to the simpler results. We will
report here on one such development, namely the construction of an event
generator for chargino production and decay which is based on the evaluation of
the full matrix elements. 

Another important feature of SUSY event generators is the possibility to impose
or relax sets of assumptions or constraints on the parameters of the model.
Several theoretical frameworks (\eg Minimal Supergravity) predict relations
between some of free SUSY parameters, and allow to produce more specific
predictions than otherwise possible. At the same time, it is however important
to be able to free themselves from relations which could artificially constrain
rates or properties of a given process, in order to make the experimental
searches as unbiased as possible. The following documentation will describe to
which extent the available codes provide such handles.

\subsection{SUSYGEN}
\begin{center}                             
\begin{tabular}{ll}
Program name:          & SUSYGEN \\
Authors:               & S. Katsanevas -- {\tt katsanevas@vxcern.cern.ch} \\
                       & S. Melachroinos -- {\tt melachr@vxcern.cern.ch} \\
Availability:          & {\tt vxcern::disk\$delphi:[katsanevas.susygen]} \\
                       & Files {\tt susygen.for} and {\tt susygen.com}\\
Documentation: & vxcern::disk\$delphi:[katsanevas.susygen]susygen\_manual.ps \\
\end{tabular}
\end{center}

SUSYGEN is a Monte Carlo generator for the production and decay of all (R-Parity
odd) MSSM sparticles in $\epem$ colliders. It is flexible enough that the user
can assume or relax different theoretical constraints, and it is easily
generalizable to extensions of the MSSM such as the Next to Minimal
Supersymmetric Standard Model (NMSSM) or R-Parity violating processes
\footnote{The parts of the code relative to Higgs and radiative decays of 
neutralinos and charginos were kindly provided by S. Ambrosanio. 
Those relative to R-Parity violation interactions by H. Dreiner.}. In
particular, R-Parity violating decays \cite{dreiner} of the \chioa\ (assumed to
be the lightest supersymmetric particle) can be selected by the user through
data cards. Each of the possible 45 R-parity violating operators described in
the New Physics Chapter of this Report is allowed. The input parameters
specifying the SUSY model are chosen to be: 
\begin{enumerate}
\item
$m_0$, the common  mass   of the  spin 0 squarks and sleptons, at the GUT 
scale.
\item $M_2$, the SU(2) gaugino 
mass parameter at the EW scale.
\item
$\mu$, the mixing parameter of the Higgs doublets at the EW scale,
\item
$\tan\beta$, the ratio of the vacuum expectation values of the
two Higgs doublets.
\item  $A$, the trilinear coupling  in the Higgs sector. This is used only for
the calculation of the third generation mixing.
\item and $m_A$, the mass of the pseudoscal Higgs. This is used only for
the calculation of the Higgs spectrum.
\end{enumerate}
Initial state radiation and an interface to JETSET  \cite{jetset} are
included.

The production and decay matrix elements are taken from ref.~\cite{bartl}.
Direct production of R-even MSSM particles, namely the neutral and charged
Higgs bosons h, H, A and ${\rm H}^\pm$, will be included in the next version of
the program.
            
Production and decay of unstable SUSY particles are factorized, and therefore
full initial/final state spin correlations are not included. Nevertheless 2-
and 3-body decays are generated using the complete matrix elements, including
contributions from all possible bosonic and fermionic intermediate states.
Decays to Higgs bosons and radiative decays of neutralinos and charginos are
included as well \cite{ambrosanio}.
Since all unstable SUSY particles are decayed before the call to JETSET, \stop\
hadronization is not included.

SUSYGEN has been tested extensively and found to agree within 1\% with ISAJET
(see next Section) in what concerns the production cross sections, and to agree
with the production and decay branching ratios generated by the code of the
authors of ref.~\cite{ambrosanio}. The code and complete documentation,
including a detailed list of cross section formulae and sample outputs from the
code, can be found in {\tt vxcern::disk\$delphi:[katsanevas.susygen]} in the
files {\tt susygen.for, susygen.com} and {\tt susygen\_manual.ps}. 
                                                                   
\noindent {\bf Decays}
Some detail on the treatment of SUSY particle decays in SUSYGEN is given here.
For the decays of the \chio`s and \chipm`s one can in general distinguish two 
regimes.
If all scalar masses are very large, or the fermions are mostly gauginos,
the decay occurs through an off-shell $W$ or $Z$ boson, {\it e.g.} 
$\chipma\rightarrow W^{*\pm}\chioa$ or
$\chiob \rightarrow Z^{*}\chioa$ and
$\chiob  \rightarrow W^{\mp}\chipma$.
In this case the BR's to the different final state leptons or quarks are mostly
determined from those of the off-shell Z and W. If instead the SUSY fermions are
mostly charginos, and some scalar lepton and/or quark has mass comparable to the
masses of $W$ and $Z$, decays mediated by the virtual scalars can dominate, and
the BR's to the corresponding fermions can be enhanced. Since it is assumed that
\chioa\ is the LSP, only two-body prompt decays of scalar particles are
considered. Should other charginos or neutralinos be lighter than a given
scalar, cascade decays through them are included. 
                    
SUSYGEN does not distinguish between three-body and two-body decays (when
e.g the decay to an on-shell scalar is posssible) since it includes the widths
of the scalars in the propagators and therefore lets the propagators force the
two-body kinematics, including all possible interferences. There is a small
region where the decays to Higgses or the radiative decays dominate: these rare
decays are included in the list of possible decays. They can be studied
separately by setting the other branching ratios to zero through the data card
DECSEL. The masses of the Higgses are calculated by using two-loop evolution
equations \cite{carena}. 
                                      
\noindent{\bf Program structure.}
SUSYGEN is divided in three stages. In the first stage the subroutine SCARDS
reads the steering cards and the subroutine SBOOK books some standard 
histograms. The standard histograms in the case of the SCAN option are:  the
masses, cross-sections and decay branching ratios in 2-d histograms of $\mu$
versus M. In the case of the no-SCAN option, the cos$\theta$ distribution of the
produced objects are reproduced. 
                                       
In the second stage the routine SUSANA initializes the masses and the branching
ratios of MSSM sparticles. The masses of sleptons and squarks are evaluated by
assuming a common $m_0$ mass at GUT unification and running it  down to
electroweak scales through Renormalization Group Equations (RGE's). Chargino and
neutralino masses and mixings are evaluated through the diagonalization of the
gaugino and Higgsino mass matrices \cite{bartl}. 

The double differential cross sections $\frac{d\sigma}{dsdt}$ have been
integrated analytically over $t$, and then integrated numerically over $s$
inside the subroutine BRANCH. Subroutine INTERF stores the results for further
generation. Particle codes are assigned by default their LUND values, while the
naming used by ISAJET 7.03 \cite{ISAJET} has been retained for comparison
purposes. 

The third  stage calculates the cross sections and generates the sparticles
requested by the user via data cards. The cross sections are computed from the
functions: 
CHARGI  (production cross section for $\chipm$),
PHOTI   (production cross section for all $\chio$),
GENSEL  (production cross section for $\sel$),
GENSELR (production cross sections for $\sell,\selr$),
GENSMUS (production cross section for $\smu,\stau,\squ $),
GENSNUE (production cross section for $\snu_e$),
GENSNU  (production cross section for $\snu$).
The user can also select through cards the luminosity available, 
so after this stage the number of events to be generated is calculated.

Unweigthed events generated according to the appropriate $\cos\theta$
distribution are produced by the routine SUSYEVE. Subroutine DECABR using the
tabulated branching ratios determines the branching ratio of the decay. SMBOD2
and SMBOD3 generate the 4-vectors of the decay products at each decay vertex.
The program loops till DECABR indicates there is no other possible decay. When
the RPARITY card is TRUE the above condition is fulfilled when we have the
lowest lying neutralino and standard particles in the products. When RPARITY is
FALSE routine LSPDECAY is called and the neutralino decays to the prescribed
standard particles. The above 4-vectors are interfaced to LUND in subroutine
SFRAGMENT where they fragment and decay. 

The last subroutines of MSSMGENE are SXWRLU which writes the LUND common block
to an external file (unit 12) and a small routine USER gives access to the LUND
common after generation. The subroutine SUSEND closes the program, and stores
the standard histograms to the file SUSYGEN.HIST. SUSYGEN uses routines from the
libraries {\tt jetset74, packlib} and {\tt genlib} and it has therefore to be
linked to them. 
\subsection{ISAJET}
\begin{center}                             
\begin{tabular}{ll}
Program name:          & ISAJET 7.16\\
Authors:               & H. Baer  -- {\tt baer@fsuhep.physics.fsu.edu} \\
                       & F. Paige, {\tt paige@bnlux1.bnl.gov"} \\ 
                       & S. Protopopescu {\tt serban@bnlux1.bnl.gov"} \\
                       & X. Tata  {\tt tata@uhhepj.phys.hawaii.edu"} \\
Availability:          & Patchy source file via anonymous ftp from \\
                       & {\tt bnlux1.bnl.gov:pub/isajet.}\\        
                       & Files: {\tt isajet.car, makefile.unix} (UNIX) and 
                         {\tt isamake.com} (VMS)\\
Documentation:         & ISAJET.DOC can be extracted from isajet.car \\
                       & via makefile.unix or isamake.com \\           
\end{tabular}
\end{center}

The program ISAJET~\cite{ISAJET}, originally developed to generate events for
hadron colliders, can also be used for event generation at $e^+e^-$
machines. In particular, the latest version, ISAJET 7.15, contains
the following SM $2\rightarrow 2$ subprocesses
\begin{eqnarray*}
e^+e^-&\to & f\bar{f},\\
e^+e^-&\to & WW,\\
e^+e^-&\to & ZZ,\\
\end{eqnarray*}
where $f=e,\mu ,\tau ,\nu_e ,\nu_{\mu},\nu_{\tau},u,d,s,c,b$ and $t$.
ISAJET includes the Fox-Wolfram final state shower QCD radiation~\cite{FW} and
Field-Feynman hadronization~\cite{FF}. Spin correlations for the 
$e^+e^-\rightarrow WW$ and $ZZ$ processes are currently neglected,
as is initial state photon radiation. ISAJET 7.15 does contain the capability
to generate events assuming longitudinally 
polarized $e^+$ or $e^-$ beams, although this
option may mainly be of interest to linear $e^+e^-$ collider enthusiasts.

ISAJET also contains a large amount of code relevant for Supersymmetry. 
Currently, one may input into ISAJET either MSSMi or SUGRA keywords,
corresponding to two different parameter sets.
For MSSM parameters, the inputs are:
\begin{eqnarray*}
MSSM1:&\ m_{\tg},\ m_{\tq},\ m_{\tell_L},\ m_{\tell_R},\ m_{\tnu},\\
MSSM2:&\ m_{\tst_L},\ m_{\tst_R},\ A_t,\ m_{\tb_R},\ A_b,  \\
MSSM3:&\ \tan\beta ,\ \mu ,\ m_A.\\
\end{eqnarray*}
The various sparticle masses and mixings are then calculated, as well as 
sparticle decay modes and branching fractions. GUT scale gaugino 
mass unification is assumed, as is the degeneracy of the first two
generations of squarks, and the first three generations of 
sleptons (although intra-generational slepton splitting is maintained).
A complete set of Higgs boson mass and coupling radiative corrections
(evaluated in the one-loop effective potential) are included, as well
as all Higgs decay modes to particles and sparticles~\cite{Higgs_i}.
An independent program ISASUSY can be extracted from ISAJET which yields
a hard copy of the various sparticle masses, parameters and
decay branching fractions.

ISAJET also can generate a sparticle spectrum given the parameter
set of the minimal supergravity (SUGRA) GUT model with radiative electroweak 
symmetry breaking~\cite{BCMPT}. In this case, the input 
parameters are:
\begin{eqnarray*}
SUGRA:\ m_0,\ m_{1/2},\ A_0,\ \tan\beta,\ sgn(\mu ).\\
\end{eqnarray*}
The top mass $m_t$ also needs to be specified. ISAJET will then
calculate sparticle masses by evolving 26 renormalization group 
equations between the weak scale and GUT scale, in an iterative
procedure, using Runge-Kutta method. Gauge coupling unification is
imposed, but not Yukawa unification. Weak scale sparticle threshold effects are
included in the gauge coupling evolution. Two loop RGE's are used for gauge
and Yukawa evolution, while one-loop RGE's are used for the other 
soft-breaking parameters. In the end, radiative electroweak symmetry
breaking is imposed, using the one-loop corrected effective potential.
A full set of radiative corrections are included for the Higgs boson masses 
and couplings. In addition, the running gluino mass is converted to a pole
gluino mass. An independent program ISASUGRA can be extracted from
ISAJET which yields a hard copy of the resultant sparticle masses, 
parameters and decay branching fractions. 

All lowest order $2\to 2$ sparticle and Higgs boson
production mechanisms have been incorporated into ISAJET. These include
the following processes~\cite{BBKMT} (neglecting bars over anti-particles):
\begin{eqnarray*}
e^+e^-&\to & \tq_L\tq_L,\ \tq_R\tq_R ,\\
e^+e^-&\to & \tl_L\tl_L,\ \tl_R\tl_R,\ \te_L\te_R ,\\
e^+e^-&\to & \tnu_{\ell}\tnu_{\ell},\\
e^+e^-&\to & \chipma\chimpa,\ \chipmb\chimpb,\ \chipma\chimpb ,\\
e^+e^-&\to & \chioi\chioj,\ (i,j=1-4),\\
e^+e^-&\to & Z h,\ Z H,\ Ah,\ AH,\ H^+ H^-.
\end{eqnarray*}
In the above, $\ell =e,\ \mu$ or $\tau$. All squarks (and also all sleptons
other than staus) are taken to be
$L$ or $R$ eigenstates, except the stops, for which $\tst_1\tst_1$,
$\tst_1\tst_2$ and $\tst_2\tst_2$ (here, $\tst_{1,2}$ being the lighter/heavier
of the top squark mass eigenstates) production is included.

Given a point in SUGRA or MSSM space, and a collider energy, 
ISAJET generates all
allowed production processes, according to their relative cross sections.
The produced sparticles or Higgs bosons are then decayed into all
kinematically accessible
channels, with branching fractions calculated within ISAJET.
The sparticle decay cascade terminates with the
lightest SUSY particle (LSP), taken to be the lightest neutralino ($\chioa$).
ISAJET currently neglects spin correlations and sparticle decay
matrix elements.
In the above reactions, spin correlation effects are only important
for chargino and neutralino pair production,
while decay matrix elements are mainly important for 3-body sparticle decays.
ISAJET 7.15 also includes capability to generate SUSY and Higgs processes
with polarized beams. Sample results from running ISAJET for LEP2 are
given in Ref.~\cite{BBMT}.

The complete card image PAM file for ISAJET 7.15 can be copied across
HEPNET, the high energy physics DECNET, from
bnlcl6::$2$dua14:[isajet.isalibrary]isajet.car.
A Unix makefile makefile.unix and a VMS isamake.com are available in the
same directory. The same files can be obtained by anonymous ftp from
bnlux1.bnl.gov:pub/isajet.                                 
                                                           
A sample input file for generating all sparticle processes at LEP2 is 
given below:
\begin{verbatim}
SAMPLE LEP2 SUGRA JOB
175.,100,0,0/
E+E-
NTRIES
2000/
SEED
999999999956781/
TMASS
180,-1,1/
SUGRA
100,80,0,2,-1/
JETTYPE1
'ALL'/
JETTYPE2
'ALL'/
END
STOP
\end{verbatim}

\subsection{SUSYXS}
\label{susyxs}
\begin{center}                             
\begin{tabular}{ll}
Program name:          & SUSYXS 1.0, Dec 15 1995\\
Authors:               & M. Mangano -- {\tt mlm@vxcern.cern.ch} \\
                       & G. Ridolfi -- {\tt ridolfi@vxcern.cern.ch} \\                           
Availability:          & {\tt http://www.ge.infn.it/LEP2} and \\
                       & {\tt http://surya11.cern.ch/users/mlm/SUSY} \\
Documentation:         & To be found in the above WWW directories \\
\end{tabular}                                                     
\end{center}
This is not an event generator, but a collection of simple programs to
evaluate total cross sections for SUSY particles in $e^+e^-$ collisions. 
No decays nor evaluation of decay BR's are included.
This set of programs is mostly useful as a reference, to obtain quickly total
production rates as a function of the various relevant parameters. It was used
during the workshop as a benchmark for the comparisons among the different
codes.  The following processes are available (each encoded in a different
fortran program):
\begin{itemize}
\item chargino pair production 
        ({\tt chargino.for}). \\
        Input parameters: $\sqrt{s}$, $M_2$, $\mu$,  
        \tanb, $M(\snu_e)$.
\item neutralino pair production, for all possible neutralino pairs
        ({\tt neutralino.for}). \\
        Input parameters: $\sqrt{s}$, $M_2$, $\mu$, 
        \tanb, $M(\sel)$.             
\item selectron pair production ($LL$, $RR$ and $RL$)
        ({\tt selectron.for}). \\
        Input parameters: $\sqrt{s}$, $M(\sel_L)$, $M(\sel_R)$, $M_2$, $\mu$, 
        \tanb,                                             
\item smuon pair production ($LL$, $RR$)
        ({\tt smuon.for}). \\
        Input parameters: $\sqrt{s}$, $M(\smu_L)$, $M(\smu_R)$.
\item stop pair production
        ({\tt stop.for}). \\
        Input parameters: $\sqrt{s}$, $M(\stop_1)$, $M(\stop_2)$, $\theta_{LR}$.
\item Higgs production
        ({\tt higgs.for}).  \\
        Input parameters: $\sqrt{s}$, $M_A$, \tanb, $M(\tilde q)$.
\end{itemize}                                                     
ISR is included, as well as QCD corrections in the case
of stop production \cite{Drees}. All references for the formulas used are
included as comments in the fortran files. The Higgs production code includes
the one-loop-corrected masses~\cite{higgssection}, using the formulas of
ref.~\cite{ERZ}. 

\noindent {\bf How the code works.}
The code relative to the process of interest has to be linked to 
{\tt phoisr} (which incorporates the ISR corrections)
and to the CERN libraries. The executable can be run interactively, and the
input parameters can be entered by the user at running time. Results with and
without ISR are printed. In the case of chargino, neutralino and higgs
production, the mass spectra are given as well. The codes are simple enough
that any user can modify them easily to customize the output and produce
directly, for example, cross section distributions or scatter plots. Likewise,
the extraction of angular distributions for most processes is straightforward, 
as all needed formulas are collected in the codes.
\subsection{SUSY23}
\begin{center}                             
\begin{tabular}{ll}
Program name:          & SUSY23 version 1.0 \\
Authors:               & J. Fujimoto, T. Ishikawa, M. Jimbo, T. Kaneko, \\ 
                       & K. Kato, S. Kawabata, T. Kon, Y. Kurihara, \\ 
                       & D. Perret-Gallix, Y. Shimizu, H. Tanaka \\
                       & {\tt susy23@minami.kek.jp} \\
Availability:          & Anonymous ftp: {\tt ftp.kek.jp} \\
                       & Files in: {\tt /kek/minami/susy23}. \\
Documentation:         & \\
\end{tabular}
\end{center}

This is a Monte-Carlo unit-weight event generator for $2\to 3$ SUSY processes
at LEP2 energies, based on the minimal supersymmetric
standard model (MSSM.).

\noindent {\bf Features of the program:}
\begin{itemize}                         
\item Processes available:
$e^+e^- \to$ 
 {\mbox{$\chi^{+}_1$}}{\mbox{$\chi^{-}_1$}}, 
 {\mbox{$\tilde{\ell}^{+}_{L,R}$}}{\mbox{$\tilde{\ell}^{-}_{L,R}$}}, 
 {\mbox{$\tilde{\nu}^{*}_{\ell}$}}{\mbox{$\tilde{\nu}_{\ell}$}},
 {\mbox{$\tilde{t}^{*}_1$}}{\mbox{$\tilde{t}_1$}}, 
  {\mbox{$\tilde{b}^{*}_1$}}{\mbox{$\tilde{b}_1$}},
 {\mbox{$\chi^{0}_1$}}{\mbox{$\chi^{0}_2$}}, 
 {\mbox{$\chi^{0}_2$}}{\mbox{$\chi^{0}_2$}}, 
 {\mbox{$\gamma$}}{\mbox{$\chi^{0}_1$}}{\mbox{$\chi^{0}_1$}}, 
 {\mbox{$e^{\pm}$}}{\mbox{$\tilde{e}^{\mp}$}}{\mbox{$\chi^{0}_1$}}, 
 {\mbox{$e^{\pm}$}}{\mbox{$\tilde{\nu}$}}{\mbox{$\chi^{\mp}_1$}}
\item Initial state radiation implemented using the structure function
approach, and using  QEDPS in some processes~\cite{qedps}
\item Final sparticle decays included (see below)
\item Hadronization realized via an interface with JETSET~\cite{jetset}.
\end{itemize}        

\noindent {\bf How the code works.}
FORTRAN source codes are generated by {\bf GRACE}~\cite{grace} which is 
a program for automatic computation of Feynman amplitudes. 
Largely exercised on standard model processes, {\bf GRACE} is being used in
the SUSY framework thanks to the addition of a dedicated
vertex and propagator library. Tools have been
developed to build automatically the SUSY23 event generator
from the various processes thus prepared. Based on an open architecture, the
generator can easily accommodate the addition of foreseen more complex
processes ($2\to4$).
The numerical integration of the differential cross section over the phase
space is carried out by the program BASES\cite{bases}.
All information on the event kinematics and the phase space hyper-cell weight
map are then used by
the event generation program SPRING\cite{bases} to produce unit-weight events.

Helicity informations are available at the parton level.
The hadronization is performed through the interface to the JETSET \cite{jetset}
package which has been extended to incorporate SUSY particle codes.

In this version (V1.0), the user may generate only one process per run, in
future releases, the possibility will be given to produce events
from a selected set of processes
accordingly to their respective probability.

\noindent {\bf Input parameters}
Two approaches have been developed to better suit the user needs:
\begin{itemize}
\item A general program contains all process codes, the selection
being performed by setting data cards.
\item An interactive tool using menus and requesters gives the user the
possibility
to build a generator dedicated to a single process.
\end{itemize}
The following parameters can be set by the user:
\begin{itemize}
\item Selection of SUSY processes
\item Center of mass energy : $\sqrt{s}$
\item Experimental cuts
    \begin{description}
    \item[$-$] angle cuts for each sparticles
    \item[$-$] energy cuts for each sparticles
    \item[$-$] invariant mass cuts
    \end{description}
\item SUSY parameters
\end{itemize}
The program is based on the MSSM and the notation for SUSY parameters
in ref.~\cite{hikasa} is adopted. 
The input SUSY parameters are: 
\begin{itemize}
\item gaugino parameters: $\tan\beta$, $M_2$, $\mu$
\item scalar lepton masses: $m_{\tilde{\ell}_L}$, $m_{\tilde{\ell}_R}$
\item scalar (light) quark masses: $m_{\tilde{q}_L}$, $m_{\tilde{q}_R}$
\item third generation scalar quark masses: 
$m_{\tilde{t}_1}$, $m_{\tilde{t}_2}$, $\theta_t$, 
$m_{\tilde{b}_1}$, $m_{\tilde{b}_2}$, $\theta_b$
\end{itemize}
General GRACE parameters can be found in the GRACE manual \cite{grace}
(Helicity amplitude techniques, diagram generation and selection, phase
space integration, event generation).

\noindent {\bf Sparticle decays.}
\label{decay}
Particle widths and decay branching ratios for all possible modes are 
calculated. Each event final state is then generated according 
to these probabilities.
We have included some possible cascade decays of sparticles as well as 
2-body and 3-body direct decays.

\noindent {\bf Check of results}
We compared the results for 2-body processes,
$e^+e^- \to$
 {\mbox{$\chi^{+}_1$}}{\mbox{$\chi^{-}_1$}},
 {\mbox{$\tilde{\ell}^{+}_{L,R}$}}{\mbox{$\tilde{\ell}^{-}_{L,R}$}},
 {\mbox{$\tilde{\bar{\nu}}_{\ell}$}}{\mbox{$\tilde{\nu}_{\ell}$}},
 {\mbox{$\tilde{\bar{t}}_1$}}{\mbox{$\tilde{t}_1$}},
 {\mbox{$\tilde{\bar{b}}_1$}}{\mbox{$\tilde{b}_1$}},
 {\mbox{$\chi^{0}_1$}}{\mbox{$\chi^{0}_2$}},
 {\mbox{$\chi^{0}_2$}}{\mbox{$\chi^{0}_2$}}
with the analytical exact calculation.
As for the 3-body processes,
$e^+e^- \to$
 {\mbox{$e^{\pm}$}}{\mbox{$\tilde{e}^{\mp}$}}{\mbox{$\chi^{0}_1$}},
 {\mbox{$e^{\pm}$}}{\mbox{$\tilde{\nu}$}}{\mbox{$\chi^{\mp}_1$}},
the results were checked against the analytical calculation based on the
equivalent photon approximation.
For the radiative process,
$e^+e^- \to$
 {\mbox{$\gamma$}}{\mbox{$\chi^{0}_1$}}{\mbox{$\chi^{0}_1$}},
we compared the result with the exact calculation for
$e^+e^- \to$
 {\mbox{$\gamma$}}{\mbox{$\tilde{\gamma}$}}{\mbox{$\tilde{\gamma}$}}
by taking a specific parameter points which corresponds to the case
{\mbox{$\chi^{0}_1$}} $\simeq$ {\mbox{$\tilde{\gamma}$}}.
The results for all 2-body processes are consistent with those of
SUSYGEN~\cite{susygen}.
\subsection{DFGT: a chargino MC generator 
with full spin correlations}
\label{DFGT}
\begin{center}                             
\begin{tabular}{ll}
Program name:          & DFGT \\
Authors:               & C. Dionisi -- {\tt dionisi@vxrm70.roma1.infn.it} \\
                       & K. Fujii -- {\tt fujiik@jlcux1.kek.jp} \\
                       & S. Giagu -- {\tt giagu@vxcern.cern.ch} \\ 
                       & T. Tsukamoto -- {\tt tsukamot@kekvax.kek.jp} \\
Availability:          & \\
Documentation:         & \\
\end{tabular}
\end{center}

\noindent {\bf General features.}
We shortly summarize the features and performances of
a new Montecarlo event generator, \DFGT \cite{cit4} which takes properly into 
account the full spin 
correlations that occur in the amplitude due to the matching 
of the spin of the produced and the decaying particle.
The choice of SUSY parameters is that of the  minimal supergravity
scenario, assuming the GUT-relations \cite{hikasa}. 
The masses and the couplings of the SUSY particles
are then specified by the four parameters \mzer, \Mdue, \Mu\ and \tanbet.              
                                                                 
\par The events are generated as follows:

\begin{itemize}
\item Full helicity amplitudes including decays into final state partons are
first calculated at tree level. This is done using \HELAS library routines 
\cite{helas}, which allows to implement correct angular correlations  and
effects of the natural widths of unstable partons.
\item The effective cross sections are then evaluated by the numerical 
integration package \BASES \cite{bases}. Initial state radiation is included
in the structure function formalism, using the results of
ref.~\cite{fujimoto}.                         
\item The generation of unweighted events is done at the partonic level  
using the \SPRING package  \cite{bases}, and the QCD evolution and
hadronization of the final state quarks in performed via an interface with
\JETSET \cite{jetset}.
\end{itemize}                                                         
Chargino pair production takes place via $s$-channel \foton\ and
\zboson\ exchange and via $t$-channel \sneutrino\ exchange.   
Only the light chargino and the 
lightest neutralino (taken as the LSP)  are currently described by \DFGT.
Furthermore, it is assumed that charginos are
lighter than all sfermions.
The case of a \sneutrino\ lighter than the chargino
\cite{cit1}, the dominant decay mode being then \chreaA, will be 
described in a forthcoming paper \cite{cit4}.             

\begin{figure}[ht]
 \begin{center}
  \mbox{\begin{tabular}[]{cc}
\subfigure[]{\epsfig{file=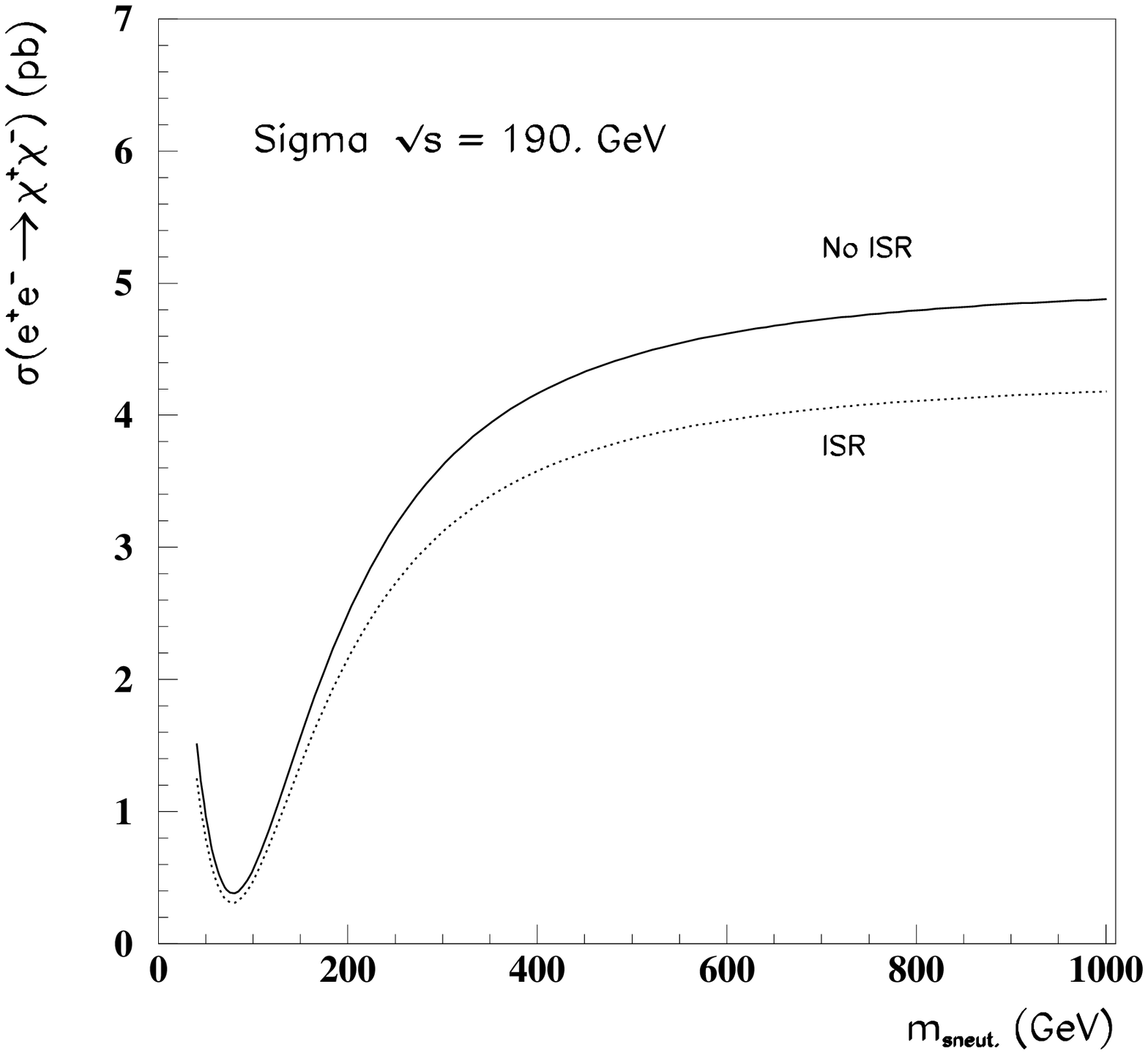,width=.40\textwidth}}  &
\subfigure[]{\epsfig{file=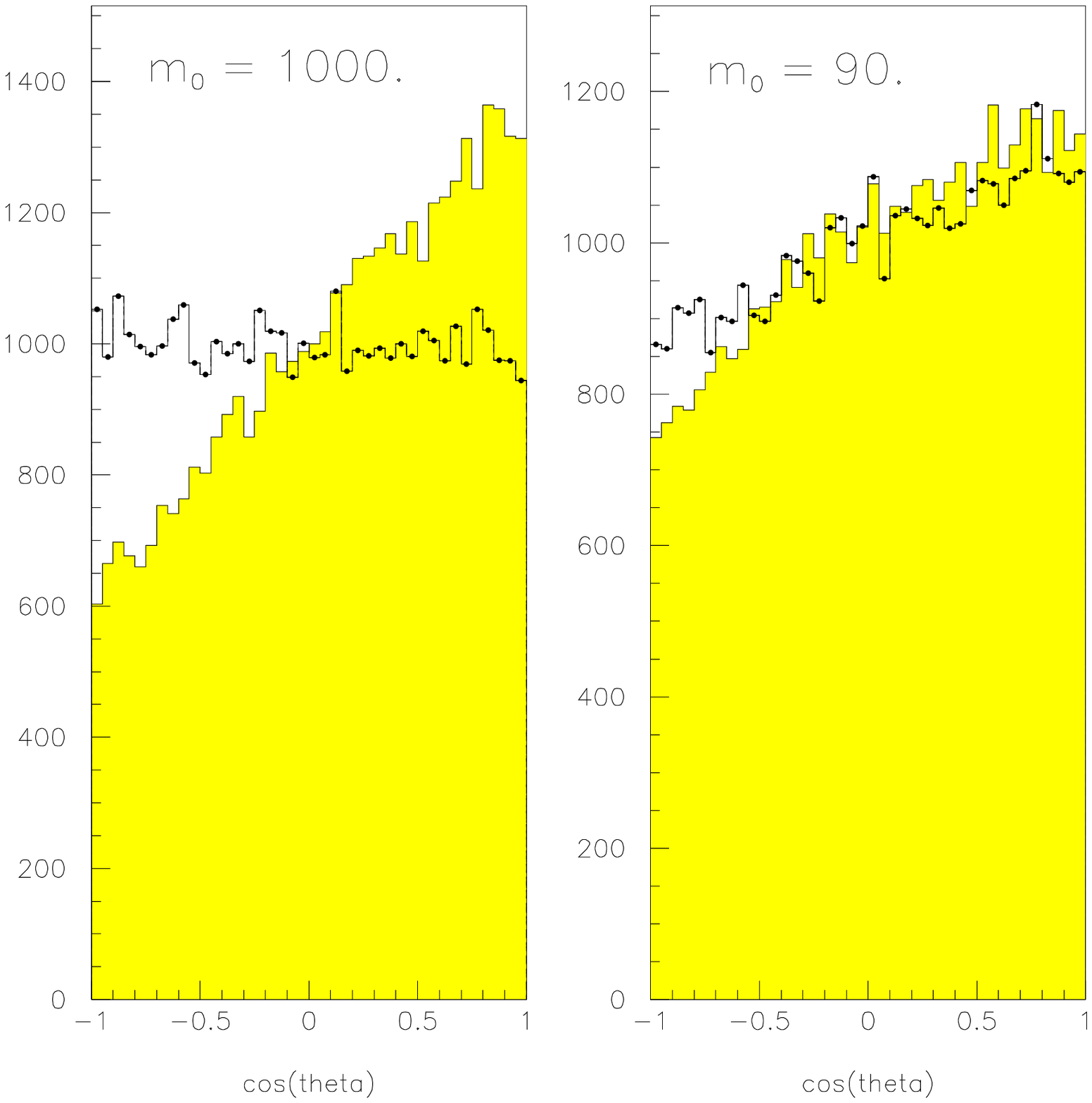,width=.40\textwidth}}
  \end{tabular}}                
 \end{center}
\ccaption{}{\label{figdfgt}                                           
(a) Total cross section for chargino pair production as function
of \Msneut and (b) angular distributions for the fermions for the set 
1A and 1B (Filled histogram: \DFGT and histogram+dots: \SUSYGEN).}
\end{figure}

\noindent {\bf DFGT performance and comparison with SUSYGEN.} 
Some results from the \DFGT Montecarlo will now be presented.
Figure~\ref{figdfgt}({\it a}) gives the total cross section of the chargino pair
production as a function of \Msneut\ showing the well known behaviour due to the
interference between the $s$-channel and the $t$-channel amplitudes. 
The total cross sections at $\sqrt{s}= 190$ GeV with and without
ISR corrections, and the total chargino widths for six points of the 
SUSY parameter space are listed in table~\ref{tabella1}.
The six points, all with $\tanbet=1.5$, correspond to the 
the following set of parameter values:
\begin{enumerate}
\item  $\Mu =-190$ GeV, $\Mdue =65$ GeV
\item  $\Mu =-180$ GeV, $\Mdue =150$ GeV
\item  $\Mu =-40$ GeV, $\Mdue =240$ GeV
\end{enumerate}
Labels A and B in table~\ref{tabella1}\ correspond to $\mzer =1000$ GeV and
$\mzer =90$ GeV, respectively.
                                 
{\renewcommand{\arraystretch}{1.3}
\begin{table}[hbtp]              
\footnotesize
\begin{center}
\begin{tabular}{|c|c|c|l|l|}
\hline                    
Set &  \multicolumn{2}{c|}{\Gcharg\ (keV)} & \multicolumn{2}{c|}{\Xsect\ (pb)} \\
\cline{2-5}                                                    
& \DFGT & \SUSYGEN & \DFGT &     SUSYXS \\ \hline
1A   &  37.69   &  37.58  &  4.849 (born) &  4.849 (born)    \\
     &          &         &  4.150 (ISR)  &  4.144 (ISR)     \\
1B   &  66.77   &  66.79  &  0.538 (born) &  0.532 (born)    \\
     &          &         &  0.452 (ISR)  &  0.449 (ISR)     \\
\hline
 2A   &  35.80   &  36.87  &  3.630 (born) &  3.623 (born)    \\
      &          &         &  3.090 (ISR)  &  3.038 (ISR)     \\
 2B   &  39.07   &  40.21  &  1.656 (born) &  1.659 (born)    \\
      &          &         &  1.415 (ISR)  &  1.419 (ISR)     \\
\hline
 3A   &   2.79   &   2.75  &  3.503 (born) &  3.551 (born)    \\
      &          &         &  3.605 (ISR)  &  3.640 (ISR)     \\
 3B   &   2.79   &   2.75  &  3.287 (born) &  3.324 (born)    \\
      &          &         &  3.393 (ISR)  &  3.419 (ISR)     \\
\hline
\end{tabular}
\end{center}
\ccaption{}{\label{tabella1}
Cross sections and total chargino widths for six points of
SUSY parameter space.}
\end{table}  }
For comparison the cross sections from SUSYXS (see section~\ref{susyxs})
and the total widths from \SUSYGEN are also given. 
The cross sections agree at the percent level, while for
the widths the agreement is of the order of few percent.
                                                
The effect of the spin correlations will now be shown by 
comparing some key distributions from \DFGT and \SUSYGEN at the 
generator level.

The angular distributions of the final state fermions for the parameter set 1A
(which gives $\Mchar =86$ GeV, $\Mneut =37$ GeV, $\Mslep \simeq \Msquar \simeq
1000$ GeV) are shown in fig.~\ref{figdfgt} ({\it a}). Here $\theta$ is the angle
between the outgoing fermion and the incoming electron. It is worth noticing
that because of the large value of \Msneut\ chargino production is dominated by
the $s$-channel contribution, with the decay mode being dominated by $\chreaB\to
f\bar f' \chio$. The peak at $\cos\theta= 1$ is entirely due to the spin
correlations, and is completely absent in the SUSYGEN distribution. 
                                               
The same distributions for the set 1B are given 
in \ref{figdfgt} ({\it b}).                                  
Contrary to case 1A, now the $t$-channel contribution to the production 
and the $\chreaC\to f\bar f' \chi0$ decay are relevant.
Although less pronounced than in \DFGT, 
the forward peak in the distribution appears now also
in the SUSYGEN case. This reflects the non-trivial chargino
angular distribution induced by the $t$-channel diagram.
More work trying to pin down in detail how spin correlations affect the angular
distributions is under way~\cite{cit4}.
                                      
\begin{figure}
 \begin{center}
  \mbox{\begin{tabular}[t]{cc}
\subfigure[]{\epsfig{file=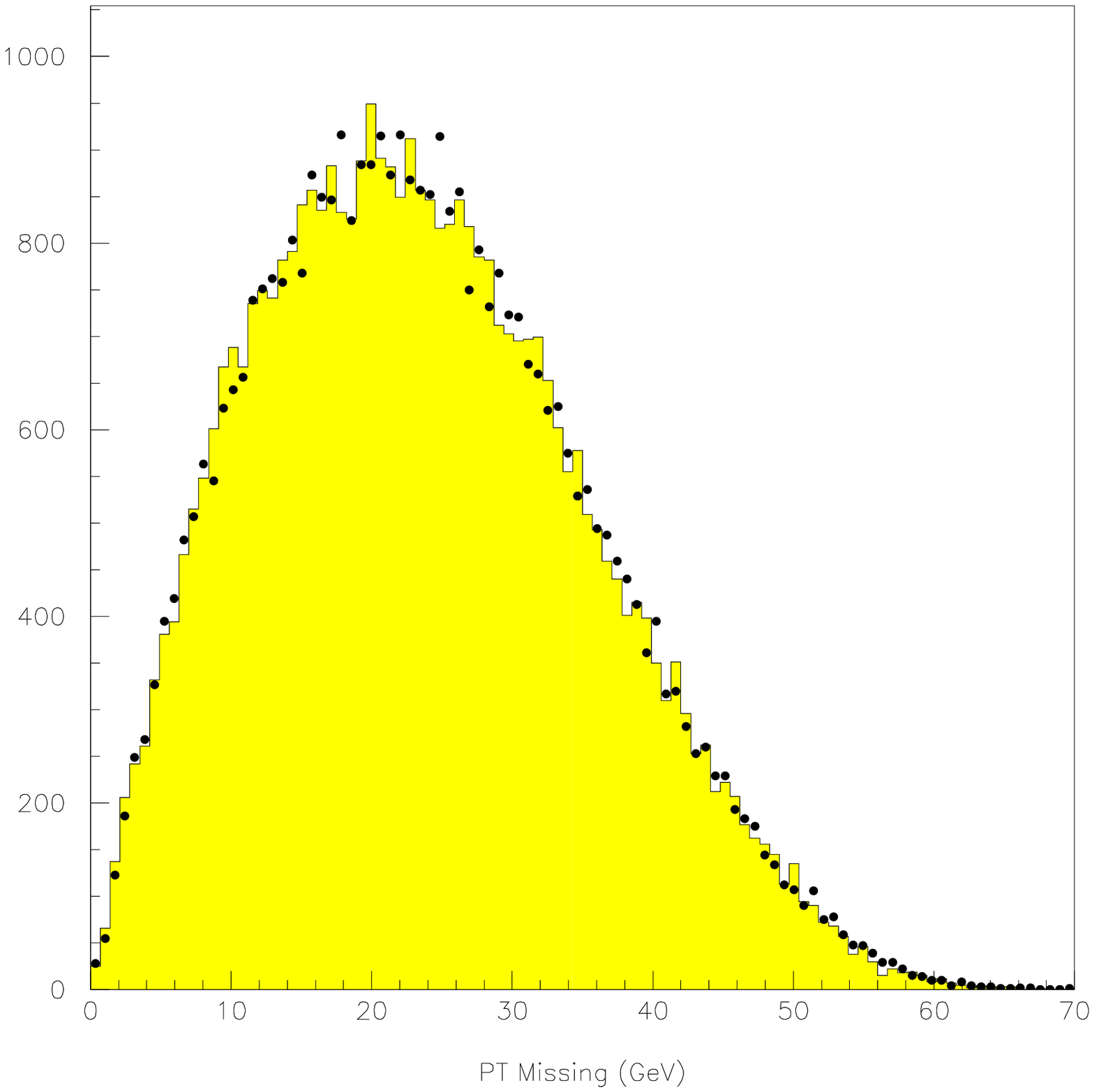,width=.40\textwidth}}  &
\subfigure[]{\epsfig{file=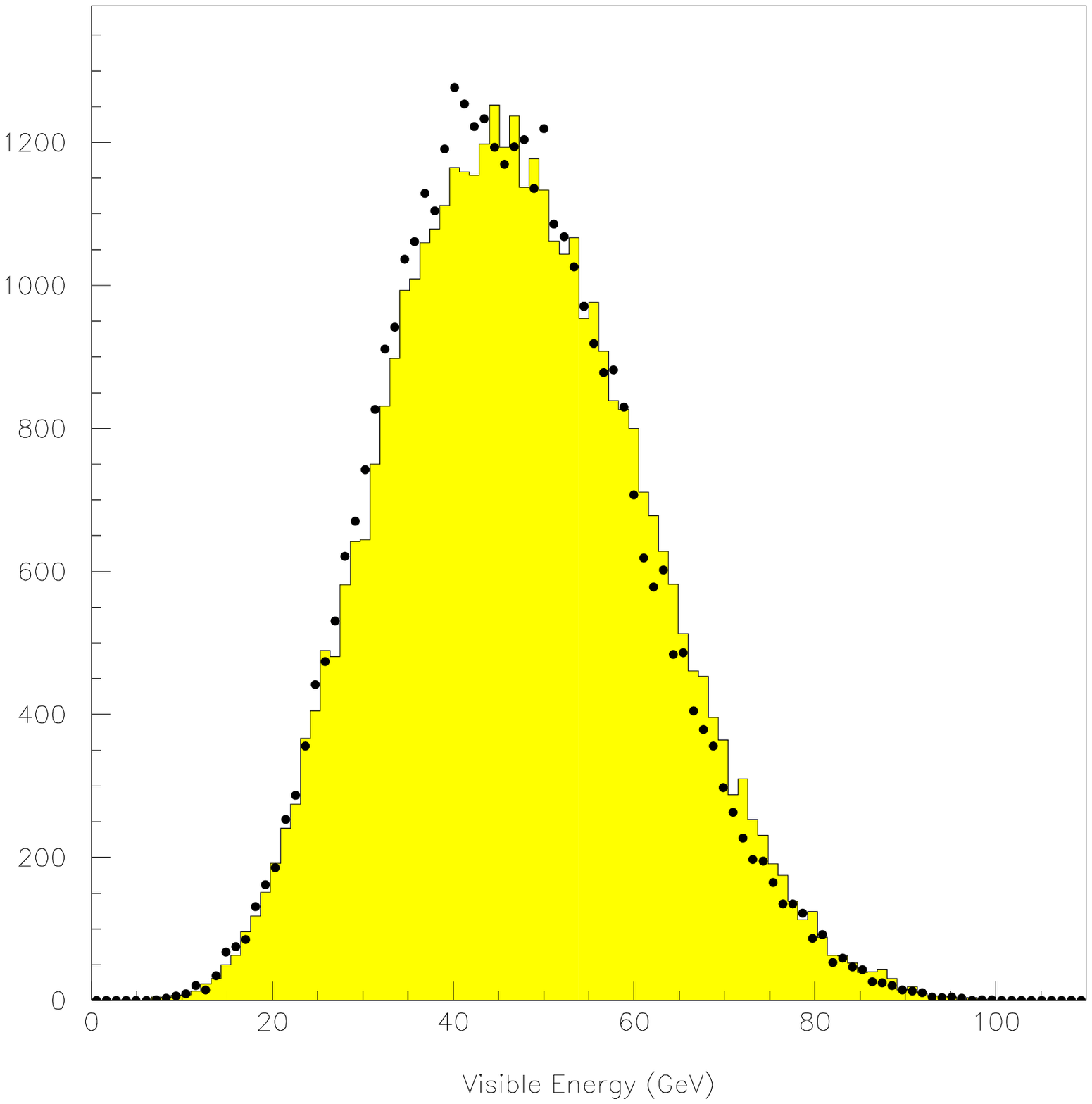,width=.40\textwidth}}
  \end{tabular}}
 \end{center}
\ccaption{}{\label{fig3}
Missing \pt (a) and Visible Energy (b) distributions,
for \DFGT (histogram) and \SUSYGEN (dots).}
\end{figure}

\begin{figure}
 \begin{center}
  \mbox{\begin{tabular}[t]{cc}
\subfigure[]{\epsfig{file=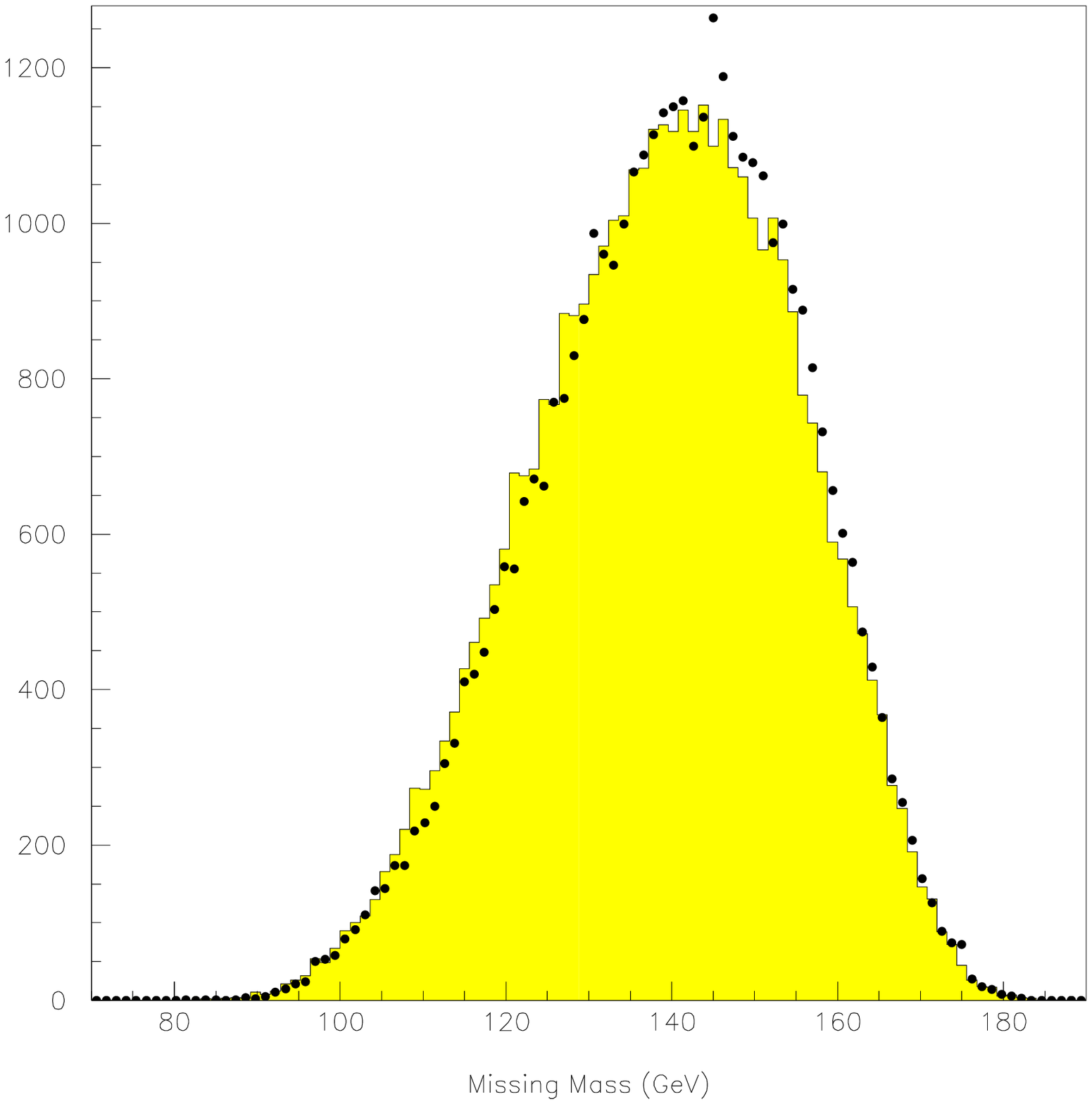,width=.40\textwidth}}  &
\subfigure[]{\epsfig{file=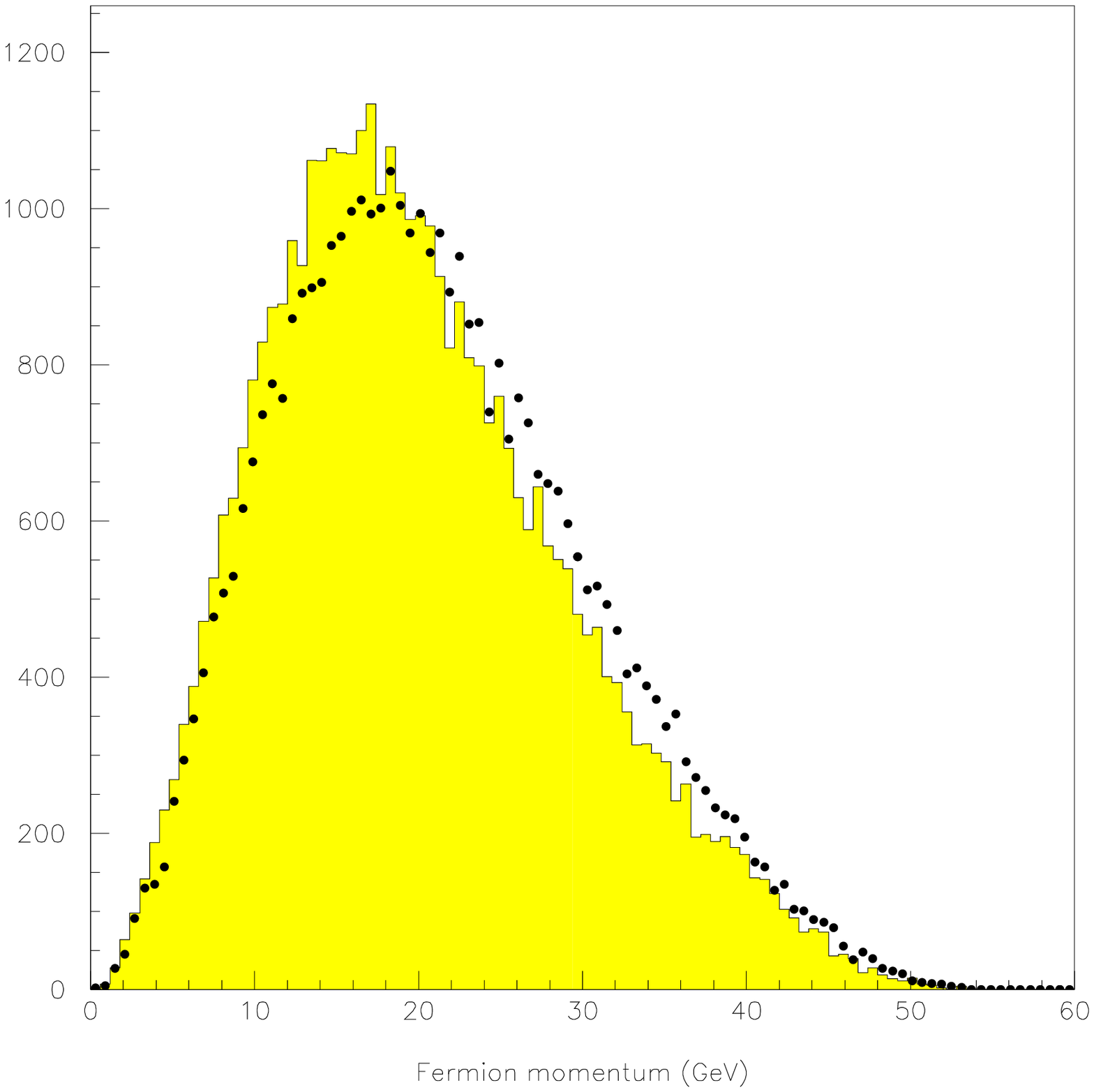,width=.40\textwidth}}
  \end{tabular}}
 \end{center}
\ccaption{}{\label{fig4}
Missing Mass (a) and Fermion Momentum (b) distributions,
for \DFGT (histogram) and \SUSYGEN (dots).}
\end{figure}

The impact of these differences on the chargino search has 
been checked by comparing at the generator level 
the distributions which play a major r\^ole in separating the signal from 
the physics backgrounds.                                  
Figs.~\ref{fig3}\ and \ref{fig4}\ show, for \DFGT and \SUSYGEN, 
the missing \pt, the visible energy, the missing mass and the
fermion-momentum distributions for set 1B (for set 1A the agreement is
very good and it is not shown here).                 
Although there is a systematic shift of about 1 GeV between the mean 
values for all the distributions, there is a good agreement in
the tails in the regions where the cuts are applied.   
The importance of such effects has also been evaluated 
through a complete analysis of the two generators and using a full L3 
detector simulation.                                        
The two analysis give essentially the same results for the sets of parameters
considered here.                                               
However it is important to point out that in other points of the SUSY 
parameter space the conclusion might be
different, in particular in regions 
where the mass splitting between the chargino 
and the neutralino is small, and where both are Higgsino-like.
These cases are currently  under investigation \cite{cit4}.    
                                                          
%
\subsection{Scalar top and scalar bottom event generators}   
The top quark has two supersymmetric partners, $\stop_L$ and $\stop_R$. The
mass eigenstates, $\stopone$ and $\stoptwo$, are mixtures of the two given by
the mixing angle $\theta_{LR}$.                
In this section 
we briefly describe and compare
the generators developed by different LEP experiments
for the production and decay 
of \stopone\ (from now on simply indicated by \stop) pairs.
As discussed in detail ref.~\cite{npreport},                        
the cross section and kinematics of the
\stop\ production is governed by two free parameters, the stop mass
\mstop\ and $\theta_{LR}$. 
The only decay channels which are of potential interest at LEP2 
are $\stop \to \chioa c$ and $\stop \to \chipa {b}$.
The latter decay channel has unit branching ratio when kinematically allowed;
otherwise, the dominant final state becomes $\chioa c$. The decay
mode is therefore completely specified by stop, chargino and neutralino masses.
The chargino then decays via $\chipa \to W^{+*} \chioa \to f \bar f' \chioa$;
the decay into a real W$^+$ is almost always forbidden in the LEP2 energy    
range. The relative values of the stop, neutralino and chargino masses are
the most significant parameters for the determination of the detection    
efficiencies.
             
The two most significant issues in the development of an event generator for
\stop\ are the treatment of the perturbative radiation off the \stop, and of
the \stop's hadronization and decay. Since the \stop\ is a scalar particle, the
spectrum of gluons emitted during the perturbative evolution will differ from
that of a quark. Therefore the standard shower evolution codes such as JETSET
would in principle require modifications in order to incorporate the correct
radiation off the \stop. The Altarelli-Parisi splitting function describing the
$\stop\to\stop g$ branching as a function of the fractional energy carried away
by the gluon ($x_g=1-x_q$) is given by: 
\be                             
        P_{\tilde q \tilde q }(x_q) = \frac{\as C_F}{2\pi} 
        \left [ \left(\frac{1+x_q^2}{1-x_q}\right)_+ - (1-x_q) \right]
\ee                     
with $C_F=4/3$, to be compared to the standard spin-1/2 case:
\be             
        P_{qq}(x_q) = \frac{\as C_F}{2\pi}
        \left(\frac{1+x_q^2}{1-x_q}\right)_+
\ee                                                                 
Notice that $P_{qq}(x) > P_{\tilde q \tilde q}(x)$,
namely the \stop\ fragmentation
function will be harder than that of a fermion of the same mass. Notice 
however that the difference is proportional to the gluon energy, and    
vanishes in the soft gluon limit ($x_q\to 1$). 
Therefore it can be consistently      
neglected within the approximations used by most shower Monte Carlo programs.
More quantitatively, one can estimate the average energy loss
due to perturbative gluon emission from a particle of mass $m$
using the well known expression \cite{khoze}:
\be                                                                           
        \langle x_g \rangle = 1- \left[\frac{\as(m)}{\as(E)} \right]
                ^{P^{(2)}/(2\pi b)},                                 
\ee                
where $P^{(2)}$ is the second moment of the relevant splitting function,
$b=(33-2N_f)/(12\pi)$ and $E$ is the beam energy. Using the values of
$P^{(2)}=-C_F$ for a spin-0 particle and $-4/3 C_F$ for spin-1/2, it is easy to
find:
\be
\langle x_g\rangle_{0} - \langle x_g \rangle_{1/2} =
 \frac{\alpha_s(E) C_F}{3\pi}\log\frac{E}{m}
\ee                         
At 190 GeV, this difference ranges between
0.01 and $3\times 10^{-3}$ for $45<\mstop<80$ GeV, with average energy losses
for the scalar case of 3\% and 1\%, respectively.
Such effects are totally negligible.                                 
                                                                       
As for the issue of \stop\ hadronization, it is important to realize that when
the dominant decay mode is $\stop\to\chioa c$ the \stop\ lifetime is longer
than the typical time scale of hadron formation, and \stop-hadrons are formed
before decay. Therefore, \stop\ hadronization must be taken into account by
Monte Carlo generator. This has been done within different approaches, which
will be described and compared in the following. 

Improvements and extentions of the existing codes, in order to achieve a more
precise description of \stop\ physics, are possible and foreseen. For details
on the individual generators, see~\cite{DELPHInote,Lthreenote,OPALnote}. 
          
\subsubsection{The DELPHI event generator.}
The DELPHI $\stop$ and $\sb_1$ event generators are based on the packages BASES
and SPRING~\cite{bases}, which perform the multidimensional phase space
integration and the event unweighting. The expression of the differential
production cross section for $\stop$ and $\sb_1$ pairs has been computed using
the results of ref.~\cite{Drees}, which include initial state QED radiation in
the collinear approximation at the leading order, and QCD corrections. 

The event generator has been interfaced with JETSET 7.3 \cite{jetset} in order
to completely describe the evolution and hadronization of the colored partons.
Perturbative gluon radiation off the \stop\ is implemented according to the
$\stop \rightarrow \stop g$ splitting function given above (see also
ref.~\cite{zerwas}), together with some additional features such as angular
ordering of the gluon shower due to soft gluon interference as described in
ref.~\cite{webe}. The formation and decay of the \stop\ hadron is then
implemented in the spectator quark approach \cite{spectator}. After the decay,
a color string is pulled between the decay $c$ quark and the spectator quark,
giving rise to the standard string fragmentation. 
                                                                          
The user can choose the values of the center of mass energy, the \stop\ mass,
the mixing angle $\theta_{\stop}$, and the $\chioa$ mass. It is also possible
to decide whether or not to include QCD corrections and/or initial state
radiation. The decay $\stop \to b \chipa$ with $\chipa \to W^{\ast} \chioa \to
f \bar f' \chioa$ is also implemented; in this case, the $\chipa$ mass is an
additional free input parameter. 

The $\sb$ event generator has been implemented along similar lines; 
the only decay mode in this case is $\sb_1 \rightarrow b \chioa$. 
                                 
\subsubsection{The L3 event generator.}
The L3 event generator~\cite{Lthreenote} includes both \chio\ and \chipm\ decay
modes. The L3 event generator is based on the calculation of 4-momenta
distributions of the final state particles $\chioa c \chioa \bar{c}$ or
$\chima b \chipa \bar{b}$. The large effects of QCD corrections are
included in the cross section calculations using the results of
ref.~\cite{vienna} (see also~\cite{zerwas}). 
The \stop\ production and decay have been defined as new
processes in PYTHIA~\cite{jetset}. The event generation process includes
modeling of hadronic final states. 
                   
In the first step of the event generation, initial state photons are emitted
using the program package REMT~\cite{jetset}, and the production cross section
at the reduced center-of-mass energy is calculated. 
The effective center-of-mass
energy is calculated for the generation of the 4-momenta of the final
state particles. These 4-momenta are then boosted according to the momentum of
the initial state photon. No perturbative gluon radiation is included before
the \stop\ decay. This is justified by the fact that less than about 1\% of the
\stop\ energy is expected to be radiated in the form of hard gluons. After the
\stop\ decay, a color string with the invariant mass of the
quark-antiquark-system (${ c} \bar{ c}$ or ${ b} \bar{ b}$) is defined. Gluon
emission and hadronization is then performed using the Lund model of string
fragmentation as implemented in PYTHIA \cite{jetset}. The Peterson
fragmentation parameters \cite{Peterson} for the $c$ and $b$-quarks are chosen
to be $\epsilon_c=0.03$ and $\epsilon_b=0.0035$, as determined from L3 event
shape distributions. Finally, short-lived particles decay into their observable
final state, where the standard L3 particle decay tables are applied. 

\subsubsection{The OPAL event generator.}
The OPAL event generator has been used by OPAL \cite{opalstop} 
in the LEP1 analyses of $\stop$ search. It only includes
the $\stop\to c\chioa$ decay. The production matrix elements are taken from
ref.~\cite{Drees,Hikasa}, including the effect of QCD corrections. In the first
step of the event generation, inital state photons are emitted taking into
account the $\stst$ cross section at the reduced center of mass energy. JETSET
\cite{jetset} is then used to perform the perturbative gluon emission. This is
done using the default emission probabilities, evaluated assuming the radiating
particle to have spin-1/2. After the perturbative evolution, Peterson
fragmentation is introduced, with the parameter $ \epsilon_{\stop} $ set to 
\begin{equation}
\epsilon_{\tilde t}=\epsilon_b
\frac{m_{ b}^{2}}{\mstop^2},
\quad \epsilon_b \ = \ 0.0057,
\ m_b \ = \ 5 \ { {\mathrm GeV}}.
\end{equation}          
As mentioned above, in the case of the $\stop\to c\chioa$ decay the $\stop$
hadronizes to form a $\stop$-hadron before it decays. $\stop$-hadrons are
therefore formed, as bound states of a $\stop$ and a light anti-quark ($\bar{
u}$, $\bar{ d}$),  $\bar{ s}$, or a diquark  (${ uu}$ etc.). As a result of the
combined perturbative and non-perturbative evolution, about 1\% [0.5\%] of the
\stop\ initial energy goes into ordinary hadrons for a 70~GeV (80~GeV) $\stop$
at $\sqrt{s} = 190$~GeV. This is consistent with the estimates given earlier. 
                                            
After the \stop\ decay, a colour string is stretched between the charm quark
and the spectator. This colour singlet system is again hadronized by JETSET.
Additional gluon bremsstrahlung is allowed in this process. The Peterson
fragmentation function is used at the end of the charm quark evolution. 

A code based on the same physical
principles was also developed by ALEPH, and has been used in their LEP1 \stop\
analysis \cite{alephstop}.

\subsubsection{Comparison of generators for $\stst$.}
We now compare some details of the OPAL, DELPHI, and L3 $\stst$ generators for
the $\chioa { c} \chioa \bar{ c}$ channel. Some differences in the features of
the final states are observed, and their origin can be found in the different
treatment of the hadronization process. In the L3 generator, \stop\ production
and decay is performed in analogy to the top quark, whose lifetime is much
shorter than the hadronization time scale. Connecting the final state $c$ and
$\bar c$ with a string implicitly assumes that the $c\bar c$ system will
radiate coherently. This is not the case for radiation whose wave-length is
smaller than the \stop\ lifetime. OPAL introduces the intermediate step of
$\stop$-hadron formation. The radiating system after \stop\ decay is then given
not by the $c\bar c$ pair, but by the two systems $c\bar q$ and $\bar c q'$,
$q$ and $q'$ being the spectator quarks. Qualitalively this will lead to lower
particle multiplicity and more collimated jets than in the L3 approach. DELPHI
introduces the emission of a large number of low energy gluons to simulate the
fragmentation of the stop bound state. In all codes, we have checked that the
effect of varying the $\epsilon$ parameter in the Peterson fragmentation
effects is very small. 
                                                                    
To illustrate the effect of the differences just mentioned,
Table~\ref{tab:nocuts} shows the total final state charged and neutral
multiplicities, as obtained from the different programs.
Table~\ref{tab:gendetails} shows multiplicities, energies and masses for
particles with a minimum energy of 500~MeV, {\it i.e.} above the typical
detector thresholds. The visible energy is essentially determined by the decay
kinematics of the $\stop$ hadron. The 3-5~GeV difference between the OPAL and
L3 generators is due to the energy of hadrons produced during the QCD evolution
of the \stop\ before it hadronizes. This difference increases for lighter
$\stop$ because of the softer fragmentation function. The particle multiplicity
found by L3 is larger than OPAL's by up to  4 charged particles per event,
depending on the \stop\ and \chio\ masses. This is consistent with what
anticipated above. The two-jet structure is expected to be clearer for events
generated by OPAL than L3 and DELPHI, because the jet evolution is localized in
the $\stop$-hadron decay. In the DELPHI generator, the cut-off for the gluon
emission is a critical parameter and may explain the larger visible energy. The
matching of the evolution scale $Q^2$ where to terminate the gluon emission
with the $Q_0^2$ parameter in the Lund QCD parton shower optimized for the Lund
string fragmentation model must be investigated for the DELPHI model. 
                                                                               
The DELPHI and L3 generators 
also include the $b \chipa \bar{b} \chima$ channel. 
A comparison between them appears in table~\ref{tab:gendetails2}.
The agreement is good, because the $\chipa$ decay is described in
a similar way in the two generators, and in both cases the hadronization takes
place in the ${ b}\bar{ b}$ system. 

The \stop-search studies are documented in~\cite{npreport}. Since the global
event signature is the large missing momentum due to the presence of two
neutralinos in the final state, the variables in the event analysis can be
chosen to be largely independent of the generator differences. Differences
related to the hadronization properties, which possibly affect the jet
structure, can be overcome by choosing different 
resolution parameters in the jet definitions.
As a net result, in spite of the differences
currently observed among these three generators the studies of the \stop\
discovery potential carried out by the three experiments are consistent with
each other~\cite{npreport}. 
\begin{table}                              
\begin{center}
\begin{tabular}{|c|c|c|c|c|}\hline
(70,50) &neutral&charged&Evis& Mvis  \\ \hline
OPAL    &   10  &  8.1  & 50 & 41  \\
DELPHI  &   22  & 19    & 56 & 46  \\
L3      &   15  & 12    & 48 & 37  \\ \hline
\end{tabular}
\end{center}
\vspace*{-0.5cm}
\ccaption{}{\label{tab:nocuts} Comparison of LEP2 generators in the
                $\chioa { c} \chioa \bar{ c}$ channel:
                neutral and charged multiplicity, visible energy (in GeV)
                and visible mass (in GeV) without a cut on the
                particle energy. Stop and neutralino masses (in GeV) are given 
                in brackets.}
\end{table}

\begin{table}
\begin{center}
\begin{tabular}{|c|c|c|c|c|}\hline
(70,50) &neutral&charged&Evis& Mvis \\ \hline
OPAL    &    6.6&  7.0  &  48&   39 \\
DELPHI  &   10  & 16    &  53&   43 \\
L3      &    9.0& 11    &  45&   35 \\
\end{tabular}

\begin{tabular}{|c|c|c|c|c|}\hline
(70,60) &neutral&charged&Evis& Mvis \\ \hline
OPAL      &    5.0&    5.8&  28&   23 \\
DELPHI    &    8.3&   14  &  35&   28 \\
L3        &    6.5&    7.7&  24&   19 \\
\end{tabular}

\begin{tabular}{|c|c|c|c|c|}\hline
(70,65) &neutral&charged&Evis& Mvis \\ \hline
OPAL      &    3.7&    4.8&  17&   14 \\
DELPHI    &    6.1&   11  &  24&   19 \\
L3        &    4.1&    5.3&  12&  9.6 \\ \hline
\end{tabular}
\end{center}
\vspace*{-0.5cm}
\ccaption{}{\label{tab:gendetails} Comparison of
                LEP2 generators in the
                $\chioa { c} \chioa \bar{ c}$
                channel: neutral and
                charged multiplicity, visible energy (in GeV) and visible
                mass (in GeV) with a minimum cut on
                the particle energy of
                500~MeV. Stop and neutralino masses (in GeV) are given 
                in brackets.}
\end{table}        
\begin{table}
\begin{center}
\begin{tabular}{|c|c|c|c|c|}\hline
(70,60,30) &neutral&charged&Evis& Mvis \\ \hline
DELPHI  &    17 &  21   & 81 &  76 \\
L3      &    15 &  20   & 79 &  74     \\ \hline
\end{tabular}
\end{center}
\vspace*{-0.5cm}
\ccaption{}{\label{tab:gendetails2} Comparison of LEP2 generators in the
                ${ b} \chipa \bar{ b} \chima$
                channel: neutral and
                charged multiplicity, visible energy (in GeV) and visible
                mass (in GeV) with a minimum cut on the particle
                energy of
                500~MeV. Stop and neutralino masses (in GeV) are given 
                in brackets.}
\end{table}

\section{Leptoquarks}
\subsection{LQ2}
\begin{center}                             
\begin{tabular}{ll}
Program name:          & \LQ\ -- Leptoquark Event Generator 1.00/04 \\
Date of last revision: & 29 September 1995 \\
Author:                & D. M. Gingrich -- {\tt gingrich@phys.ualberta.ca}\\
Other programs called: & JETSET 7.405 (plus PYTHIA 5.710) \\
                       & and CERNLIB (DIVON4, RANECU) \\
Comments:              & source code managed with CMZ\\
Availability:          & The complete code documentation is available 
                         from the author

\end{tabular}
\end{center}
This section describes a Monte Carlo program which generates 
pair production of scalar or vector leptoquarks in electron-positron
annihilation.
The leptoquarks are produced according to an effective Lagrangian
with the following properties~\cite{br}: 1) baryon and lepton number
conservation, 2) non-derivative and family diagonal couplings to
lepton-quark pairs and 3) $SU(3)_C\times SU(2)_L\times U(1)_Y$
invariance.

The contributions to leptoquark pair production from the $s$-channel
exchange of an electroweak boson, $t$-channel exchange of a quark and
the interference between them are included in the differential
cross-section.
The angular distribution of the scalar or vector leptoquarks assumes
unpolarized beams. Initial state radiation, currently not present,
will soon be included.
The centre of mass energy is not restricted to the $Z$-resonance.
The leptoquarks are allowed to decay to lepton-quark or
neutrino-quark final states.
Decays to all three generations are possible but the massless quark
approximation will not be valid for decays to the top quark.

The LUND routines of JETSET~\cite{jetset} are used for the final state
parton shower, fragmentation and decay processes.
The generator fills the JETSET common block /LUJETS/ and the standard
Monte Carlo generator common block /HEPEVT/.
The mechanics of the program closely follows that of an analogous
generator for simulating leptoquark production and decay in
electron-proton collisions~\cite{lquark}.

\noindent {\bf Physics Processes.}
The lowest order Feynman diagrams for leptoquark production in
electron-positron annihilation ($\rm e^+e^-\ra L_Q\overline{L_Q}$) are
straightforward to evaluate using the general couplings from the
effective Lagrangian~\cite{br}.
In general, the pair production amplitudes for the s-channel and
t-channel processes can interfere and the differential cross-section
for the production of scalar leptoquarks is given by three terms:
\begin{equation}
\frac {d\sigma_{\rm scalar}} {d(\cos\theta)} = \frac {3\pi\alpha^2}
{8s} \beta^3\sin^2\theta \sum_{a=\rm L,R} [ |A_\gamma+A_{\rm Z}|_a^2 +
2\lambda_a^2{\sl Re}[(A_\gamma+A_{\rm Z})_a(A_{\rm q}^*)_a] +
\lambda_a^4|A_{\rm q}|_a^2],
\end{equation}
where $A_\gamma$ and $A_{\rm Z}$ denote the photon and Z-boson
s-channel exchange terms, and $A_{\rm q}$ is the t-channel exchange
term.
The sum is over electron polarizations and $\lambda_{\rm L,R}$ are the
generalized couplings.
$\theta$ is the polar angle and $\beta = \sqrt{1-4m_{\rm LQ}^2/s}$ is
a kinematic threshold factor.

Similarly the differential cross-section for the production of vector
leptoquarks is
\begin{equation}
\frac {d\sigma_{\rm vector}} {d(\cos\theta)} = \frac {3\pi\alpha^2}
{8M_{LQ}^2} \beta^3 \left( \frac {7-3\beta^2}{4} \right) \sum_{a=\rm L,R}
[ |A_\gamma+A_{\rm Z}|_a^2 + 2\lambda_a^2{\sl Re}[(A_\gamma+A_{\rm
Z})_a(A_{\rm q}^*)_a] +  \lambda_a^4|A_{\rm q}|_a^2 ],
\end{equation}

\begin{table}[ht]
\begin{center} 
\begin{tabular}{||c|c|c|c||} \hline
{\em $^Q${\sc Lq}$_T$} & {\em $T_3$} & {\em Decay} & {\em Coupling} \\ \hline\hline
$\rm ^{-1/3}S_0$             & $0$    & $\rm e_L^-u_L$      & $\rm  \lambda_L$ \\
$\rm ^{-1/3}S_0$             & $0$    & $\rm e_R^-u_R$      & $\rm  \lambda_R$ \\
$\rm ^{-1/3}S_0$             & $0$    & $\rm \nu_ed_L$      & $\rm -\lambda_L$ \\ \hline
$\rm ^{-4/3}\tilde{S}_0$     & $0$    & $\rm e_R^-d_R$      & $\rm  \lambda_R$ \\ \hline
$\rm ^{+2/3}S_1$             & $+1$   & $\rm \nu_eu_L$      & $\rm  \sqrt{2}\lambda_L$ \\
$\rm ^{-1/3}S_1$             & $0$    & $\rm \nu_ed_L$      & $\rm -\lambda_L$ \\
$\rm ^{-1/3}S_1$             & $0$    & $\rm e_L^-u_L$      & $\rm -\lambda_L$ \\
$\rm ^{-4/3}S_1$             & $-1$   & $\rm e_L^-d_L$      & $\rm -\sqrt{2}\lambda_L$ \\ \hline
$\rm ^{-1/3}V_{1/2}$         & $+1/2$ & $\rm \nu_ed_R$      & $\rm  \lambda_L$ \\
$\rm ^{-1/3}V_{1/2}$         & $+1/2$ & $\rm e_R^-u_L$      & $\rm  \lambda_R$ \\
$\rm ^{-4/3}V_{1/2}$         & $-1/2$ & $\rm e_L^-d_R$      & $\rm  \lambda_L$ \\
$\rm ^{-4/3}V_{1/2}$         & $-1/2$ & $\rm e_R^-d_L$      & $\rm  \lambda_R$ \\ \hline
$\rm ^{+2/3}\tilde{V}_{1/2}$ & $+1/2$ & $\rm \nu_eu_R$      & $\rm  \lambda_L$ \\
$\rm ^{-1/3}\tilde{V}_{1/2}$ & $-1/2$ & $\rm e_L^-u_R$      & $\rm  \lambda_L$ \\ \hline
$\rm ^{-2/3}V_0$             & $0$    & $\rm e_L^-\bao{d}_R$ & $\rm  \lambda_L$ \\
$\rm ^{-2/3}V_0$             & $0$    & $\rm e_R^-\bao{d}_L$ & $\rm  \lambda_R$ \\
$\rm ^{-2/3}V_0$             & $0$    & $\rm \nu_e\bao{u}_R$ & $\rm  \lambda_L$ \\ \hline
$\rm ^{-5/3}\tilde{V}_0$     & $0$    & $\rm e_R^-\bao{u}_L$ & $\rm  \lambda_R$ \\ \hline
$\rm ^{+1/3}V_1$             & $+1$   & $\rm \nu_e\bao{d}_R$ & $\rm  \sqrt{2}\lambda_L$ \\
$\rm ^{-2/3}V_1$             & $0$    & $\rm e_L^-\bao{d}_R$ & $\rm -\lambda_L$ \\
$\rm ^{-2/3}V_1$             & $0$    & $\rm \nu_e\bao{u}_R$ & $\rm  \lambda_L$ \\
$\rm ^{-5/3}V_1$             & $-1$   & $\rm e_L^-\bao{u}_R$ & $\rm  \sqrt{2}\lambda_L$ \\ \hline
$\rm ^{-2/3}S_{1/2}$         & $+1/2$ & $\rm \nu_e\bao{u}_L$ & $\rm  \lambda_L$ \\
$\rm ^{-2/3}S_{1/2}$         & $+1/2$ & $\rm e_R^-\bao{d}_R$ & $\rm -\lambda_R$ \\
$\rm ^{-5/3}S_{1/2}$         & $-1/2$ & $\rm e_L^-\bao{u}_L$ & $\rm  \lambda_L$ \\
$\rm ^{-5/3}S_{1/2}$         & $-1/2$ & $\rm e_R^-\bao{u}_R$ & $\rm  \lambda_R$ \\ \hline
$\rm ^{+1/3}\tilde{S}_{1/2}$ & $+1/2$ & $\rm \nu_e\bao{d}_L$ & $\rm  \lambda_L$ \\
$\rm ^{-2/3}\tilde{S}_{1/2}$ & $-1/2$ & $\rm e_L^-\bao{d}_L$ & $\rm  \lambda_L$ \\ \hline
\end{tabular}
\ccaption{}{\label{lqtable} Quantum numbers ($Q$ is the electric charge, $T$
is the weak isospin and $T_3$ is the third component of isospin),
coupling constants and decay channels for leptoquarks.}
\end{center}
\end{table}

From the effective lagrangian one can obtain the various partial
leptoquark decay widths, $\Gamma_{\rm LQ}$.
For the scalar (S) and vector (V) leptoquarks we have
\begin{equation}
\Gamma^{\rm S}_{\rm LQ} = \frac {\lambda^2_{\rm L,R} m_{\rm LQ}}
{16\pi} \ \ \ \ {\rm and} \ \ \ \ \Gamma^{\rm V}_{\rm LQ} = \frac
{\lambda^2_{\rm L,R} m_{\rm LQ}} {24\pi},
\end{equation}
where $\lambda_{\rm L,R}$ denote the leptoquark couplings
to a particular final state and $m_{\rm LQ}$ is the
leptoquark mass.
The total widths are obtained by summing over all possible
final states.

Table~\ref{lqtable} gives the quantum numbers, couplings and decay channels for
all leptoquarks.              
We have adapted the notation of ref.~\cite{AACHEN}\footnote{$\rm S_0,
\tilde{S}_0, S_1, V_{1/2}, \tilde{V}_{1/2}, V_0, \tilde{V}_0, V_1,
S_{1/2}, \tilde{S}_{1/2}$ in ref.~\cite{AACHEN} correspond to $\rm S_1,
\tilde{S}_1, S_3, V_2, \tilde{V}_2, U_1, \tilde{U}_1, U_3, R_2,
\tilde{R}_2$ in ref.~\cite{br} respectively.}.

\noindent {\bf Generator.}
The user must supply his own main program to initialize the package and
generate events. The initialization routine LQINIT must be called once to
perform some initialization and calculate the total cross-section. Some simple
checks are make to see that the required leptoquark and decay process are
consistent with the requested quantum numbers and couplings. A call is
automatically made to the routine TOTSCALAR or TOTVECTOR to calculate the total
cross-section. The differential cross-section function XSCALAR or XVECTOR is
numerically integrated as a test that the generator is initialized properly.
The resonance width and branching ratio are also calculated. A program banner,
the value of some parameters and the process to be generated are printed out. 

Events are generated by calling the routine LQGEN once per event in the user
main program. The routine to create the event record, LQFILL, is then
automatically called by LQGEN. Routines from JETSET are used for final state
fragmentation and decay processes. 

All other subroutines and functions are called internally. But, if so desired,
the total cross-section functions or differential cross-section functions
(function of polar angle) may be called by the user after initialization. 

\noindent {\bf Numerical integration.}
The differential cross-sections are integrated and sampled using the CERNLIB
package DIVON4~\cite{DIVON4}. The package consists of a collection of routines
to aid in the numerical integration of functions of several variables and to
sample points in a multi-dimensional coordinate space from a specified
probability density function. The algorithm adaptively partitions a
multi-dimensional coordinate space into a set of axis-oriented
hyper-rectangular regions, based on a user provided function. These regions are
then used for a stratified sampling estimate of the integral of the function,
or to sample random vectors from the coordinate space with probability density
that of the function. 
The integration and importance sampling are extremely fast in \LQ\
since the cross-section is a function of a single variable.

{\bf Installation and availability.}
The \LQ\ package is managed as a CMZ library. The program needs to be linked
with JETSET version 7.4 and PYTHIA version 5.7. The CERN libraries MATHLIB and
KERNLIB must also be loaded to include the random number generator RANECU 
timing routine TIMED and the integration package DIVON4. 

The \LQ\ CMZ library can be obtained via anonymous ftp at
{\tt jever.phys.ualberta.ca} in file {\tt pub/lq2.cmz}.

\end{document}